\documentclass[tradiabstract]{aa} 
\usepackage{graphicx}
\usepackage{supertabular}
\usepackage{txfonts}
\begin{document}
\title{Binary Planetary Nebulae Nuclei towards the Galactic Bulge}
   \subtitle{I. Sample Discovery, Period Distribution and Binary Fraction}

   \author{B. Miszalski
           \inst{1,2}
           \and
           A. Acker
           \inst{1}
           \and
           A. F. J. Moffat
           \inst{3}
           \and
           Q. A. Parker
           \inst{2,4}
           \and
           A. Udalski
           \inst{5}
          }
\institute{Observatoire Astronomique, Universit\'e de Strasbourg, 67000, Strasbourg, France\\
\email{brent@newb6.u-strasbg.fr, acker@newb6.u-strasbg.fr}
         \and
Department of Physics, Macquarie University, Sydney, NSW 2109, Australia\\ 
             \email{brent@ics.mq.edu.au, qap@ics.mq.edu.au}
         \and
         D\'ept. de physique, Univ. de Montr\'eal C.P. 6128, Succ. Centre-Ville, Montr\'eal, QC H3C 3J7, and Centre de recherche en astrophysique du Qu\'ebec, Canada\\
         \email{moffat@astro.umontreal.ca}
         \and
         Anglo-Australian Observatory, Epping, NSW 1710, Australia
         \and
         Warsaw University Observatory, Al. Ujazdowskie 4, PL-00-478, Warsaw, Poland\\
         \email{udalski@astrouw.edu.pl}
         }
   \date{Received -; accepted -}

 
   \abstract{
      Binarity has been hypothesised to play an important, if not ubiquitous, role in the formation of planetary nebulae (PNe).
      Yet there remains a severe paucity of known binary central stars required to test the binary hypothesis and to place strong constraints on the physics of the common-envelope (CE) phase of binary stellar evolution. 
      Large photometric surveys offer an unrivalled opportunity to efficiently discover many binary central stars. 
      We have combined photometry from the OGLE microlensing survey with the largest sample of PNe towards the Galactic Bulge to systematically search for new binaries. A total of 21 periodic binaries were found thereby more than doubling the known sample. 
      The orbital period distribution was found to be best described by CE population synthesis models when no correlation between primary and secondary masses is assumed for the initial mass ratio distribution. A comparison with post-CE white dwarf binaries indicates both distributions are representative of the true post-CE period distribution with most binaries exhibiting periods less than one day.
      An estimated close binary fraction of 12--21\% is derived and is the first robust and independent validation of the prevailing 10--15\% fraction estimated by Bond (2000). This suggests that binarity is not a precondition for the formation of PNe and that close binaries do not play a dominant role in the shaping of nebular morphologies. Systematic effects and biases of the survey are discussed with implications for future photometric surveys.
   }
   \maketitle
\section{Introduction}
The common-envelope (CE) phase of binary stellar evolution describes a close binary immersed within a large envelope of material lost from the primary component (Iben \& Livio 1993). It is established after mass transferred to the secondary overflows the Roche lobe surrounding the secondary and engulfs the binary system. Energy and angular momentum are transferred from the binary system to the CE and the components are forced closer together. Evidence for the CE phase is seen in a diverse range of short orbital period phenomena that include cataclysmic variables, subdwarf B or white dwarf binaries, low-mass X-ray binaries, novae, degenerate binaries such as type Ia supernovae progenitors and the close binary central stars of planetary nebulae (CSPN). 

Despite its crucial role in binary stellar evolution, there are critical aspects of the CE phase that are not well understood. This is exemplified by considerable uncertainty in the value of the $\alpha_{CE}$ parameter that measures the conversion efficiency of orbital energy required to eject the envelope. Numerous population synthesis models have explored the role of $\alpha_{CE}$ albeit with varying model assumptions and definitions of $\alpha_{CE}$ (de Kool 1990, 1992; de Kool \& Ritter 1993; Yungelson, Tutukov \& Livio 1993; Han et al. 1995; Politano \& Weiler 2007). 
The models predict observed period distributions of post-CE binaries that are very sensitive to the assumed values of $\alpha_{CE}$ and the initial mass ratio distribution (IMRD) of binaries before the CE phase. 
The IMRD itself is a highly contentious field of study subject to many selection effects (Trimble 1990; Goldberg, Mazeh \& Latham 2003; Halbwachs et al. 2003). 
Comparison of the period distributions of known post-CE binaries therefore offers a valuable avenue to constrain $\alpha_{CE}$ and the IMRD. 

The short $\sim$10$^4$ year lifetime of planetary nebulae (PNe) means that \emph{close binary CSPN are ideal probes of CE evolution} because the orbital period distribution is observed \emph{directly at the end of the CE phase} where there has been no chance for significant orbital evolution to take place as in older systems like cataclysmic variables. However, this potential has yet to be fully realised due to the previously very small sample size of binary CSPN.

De Marco, Hillwig \& Smith (2008, hereafter DM08) summarise the known close binary CSPN population of around a dozen objects. The rather inhomogeneous sample results mainly from the photometric monitoring of $\sim$100 CSPN over 30 years after many different observing campaigns by Bond and collaborators (Bond et al. 2000 and ref. therein, hereafter B00). The survey biases and characteristics are therefore not well understood (DM08) and these are compounded by a somewhat necessary bias towards low surface brightness PNe to reduce the effects of nebular contamination. The very small sample size and reservations concerning the survey biases have hindered comparisons with predicted period distributions required to constrain $\alpha_{CE}$ or the IMRD. 

B00 estimate a close binary fraction of 10--15\%, however DM08 expressed concerns about the accuracy of this very important quantity because of the survey biases. The close binary fraction is incredibly important for two interdependent reasons: (i) To investigate theoretical claims that PNe form largely, if not entirely, as a result of binary evolution (Paczynski 1985; Moe \& De Marco 2006), and (ii) To determine the degree of influence binarity has in shaping the varied and complex nebular morphologies of PNe (e.g. Balick \& Frank 2002; Soker 1997). The latter topic has been the subject of considerable debate for the last 30 years. Binarity offers a simpler or even preferred mechanism to explain non-spherical nebular morphologies such as bipolar PNe (e.g. Soker 1998), but without a firm estimate of the binary fraction other mechanisms cannot be ruled out (for a review of these other mechanisms see Balick \& Frank 2002).

New surveys for close binary CSPN are therefore required to make substantial progress in these areas by properly characterising the population and obtaining an independent estimate of the binary fraction. Ideally a photometric and radial velocity (RV) survey of all visible central stars of the $\sim$200 local volume PNe identified within 1--2 kpc (Frew 2008) would be conducted. However, this is difficult work for such a sparsely distributed sample and survey biases may be difficult to control (e.g. sampling and weather conditions). The most efficient and productive, albeit magnitude limited, would be a photometric survey of a large population of PNe either towards the Galactic Bulge or Magellanic Clouds. Fortunately, the multi-epoch photometry for such an endeavour already exists thanks to microlensing surveys dedicated to these regions.

After comprehensive exploration of the SuperCOSMOS H$\alpha$ Survey (SHS, Parker et al. 2005) the Macquarie/AAO/Strasbourg H$\alpha$ PNe catalogues (Parker et al. 2006; Miszalski et al. 2008a) have more than doubled the number of PNe towards the Galactic Bulge. We have combined these new discoveries with already catalogued PNe to search for new binary CSPN towards the Bulge using photometry with excellent spatial and temporal coverage from OGLE-III (OGLE; Sec. \ref{sec:ogle}). This survey dwarfs all previous efforts by examining nearly 300 PNe in an efficient and relatively uniform manner. Miszalski et al. 2008b presented some initial findings of the survey which we describe in full here.

Section \ref{sec:obs} introduces the observational data used during the search method described in Sec. \ref{sec:search}. We present the new binaries and compare the period distribution with population synthesis model predictions in Sec. \ref{sec:results}. An estimate of the binary fraction is also made in Sec. \ref{sec:results} and we conclude in Sec. \ref{sec:conclusion}.

\section{Observations}
\label{sec:obs}
\subsection{OGLE photometry}
\label{sec:ogle}
\begin{figure*}
   \centering
      \includegraphics[scale=0.6,angle=270]{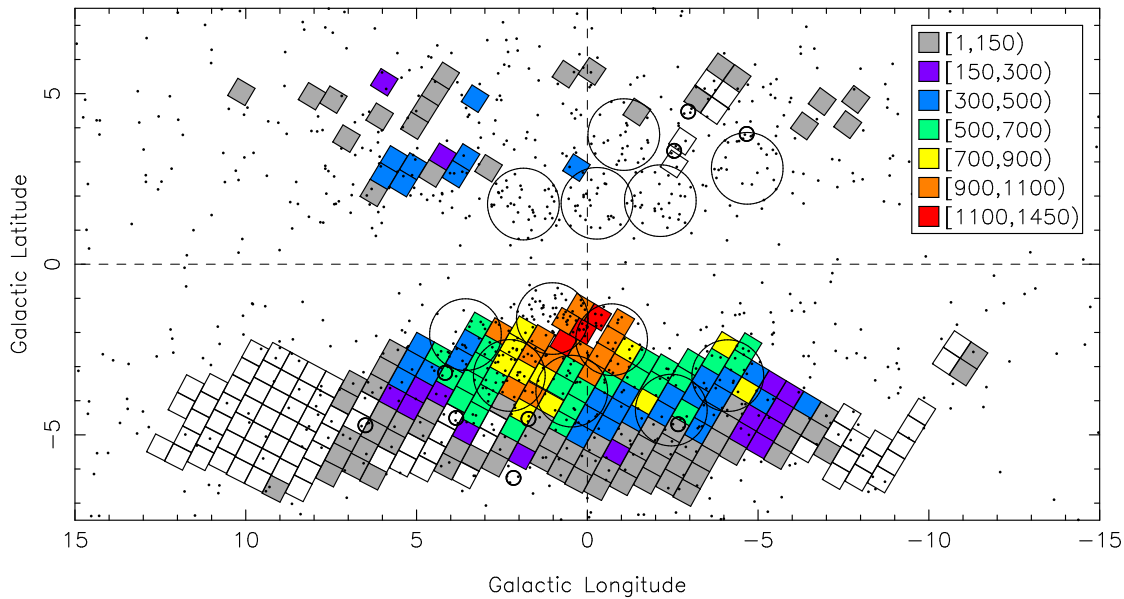}\\
      \caption{Galactic distribution of 35.5 $\times$ 35 arcmin$^2$ OGLE-III fields (squares) and PNe (points). Fields are coloured based on the number of total observations $N$, with warmer colours indicative of greater $N$ (see key for $N$ range). AAOmega fields (large circles) and FLAMES fields (small circles) indicate regions with relatively optical deep spectroscopy. 
}
   \label{fig:survey}
\end{figure*}

The Optical Gravitational Lensing Experiment (OGLE) constitutes an extensive $I$-band photometric database constructed over many years of monitoring the Galactic Bulge and Magellanic Clouds. OGLE-I, being the first phase of the experiment, monitored a few select fields towards the Bulge (Udalski et al. 1992). The second phase, OGLE-II, brought much improved temporal and spatial coverage of the Galactic Bulge (Udalski et al. 2002b;Wo\'zniak et al. 2002), but it was not until the third and most recent phase, OGLE-III (Udalski et al. 2002a, 2008) that OGLE photometry became suitable to uniformly assess CSPN binarity for a large population of PNe. 
OGLE-III is very well suited to the task with both large spatial coverage closely matching the spatial distribution of PNe (Fig. \ref{fig:survey}), and well sampled fields sensitive to detecting the short periods of binary CSPN. 
Indeed, OGLE-III is effectively  a `one-stop shop' that supersedes previous piecemeal and biased photometric CSPN monitoring campaigns from which any binary fraction estimates are rather uncertain. We therefore focus almost exclusively on OGLE-III in this work, but we have made some use of OGLE-II data in a reappraisal of Miszalski et al. (2009) to reveal one new binary (see later). 

The OGLE-III survey uses the 1.3-m Warsaw telescope at the Las Campanas Observatory, Chile. It is equipped with a CCD mosaic camera with eight 2K x 4K CCDs giving a 35 $\times$ 35.5 arcmin$^2$ field of view with 0.26\arcsec\, pixels sampling a median seeing of 1.2\arcsec\, (at worst 1.8\arcsec). There are 267 survey fields tiled such that regions of highest stellar density and lowest reddening are favoured, but 80 of these have not yet been observed (Fig. \ref{fig:survey}). Fields with higher stellar density are sampled more frequently.

A non-standard $I$-band filter is used with a bandpass that extends past the standard $\sim$9000 \AA\, drop-off (Udalski et al. 2002b) and reaches a limiting magnitude of $I\sim20$. Nebular contamination can occur for PNe with strong [SIII] $\lambda$9069, $\lambda$9532 emission lines that can be bright even in very reddened Bulge PNe (e.g. Jacoby \& Van de Steene 2004). Other prominent lines such as [ArIII] 7750, [Cl IV] 8045, HeII 8237 and members of the Paschen series lie favourably within the filter band-pass and could make a significant contribution if the reddening is not too high. A strong nebular continuum may also influence the $I$-band image in the brightest PNe. 

The vast amounts of OGLE-III data are processed by a data reduction pipeline based on the difference image analysis (DIA) method (Alard \& Lupton 1998; Wo\'zniak 2000; Udalski 2003). DIA essentially measures intensity differences between an observed frame and a high quality reference image to produce better quality time series photometry with lower scatter than other methods (Wo\'zniak 2000). Systematic effects are generally very small but can become problematic for PNe where nebular contamination is high (Section \ref{sec:systematic}). 

\subsection{Optical Spectroscopy}
\label{sec:spec}
To complement OGLE-III photometry we have procured optical spectroscopy for a large number of PNe towards the Bulge with the 2dF/AAOmega and FLAMES multi-object spectroscopy facilities. Our extensive spectroscopic data goes deeper than comparable work conducted on 2-m class telescopes (Cuisinier et al. 2000; G\'orny et al. 2004; Exter, Barlow \& Walton 2004) enabling detection of many faint diagnostic features which we use to eliminate contaminants from our PNe sample (Sec. \ref{sec:mimic}). Figure \ref{fig:survey} depicts the spatial coverage of AAOmega (large circles) and FLAMES (small circles) fields that overlap a large fraction of the better sampled OGLE-III fields. Additional spectroscopy is sourced from the MASH project (Parker et al. 2006; Miszalski et al. 2008a). The spectroscopic data are described below.

The 2dF multi-object spectroscopy facility on the 4-m Anglo-Australian Telescope enables 400 objects to be observed simultaneously over a 2-degree field of view using optical fibres (Lewis et al. 2002). The fibres are positioned in the focal plane robotically and feed the AAOmega optical spectrograph (Sharp et al. 2006). A total of 12 AAOmega fields were observed on our behalf in service mode on 26--27 March 2007 and 16 March, 29--30 May and 8 August 2008. We used the 580V and 385R volume-phase holographic gratings to give central resolutions of 3.5 \AA\, and 5.3 \AA\, (FWHM) in the blue and red spectrograph arms, respectively.
The standard dichroic was used to provide a wavelength coverage of 3700--8850 \AA, however the August field ($\ell=-2.5$, $b=-4.3$) was observed with the redder dichroic which covers 4800--9740 \AA. Exposure times ranged from 2--3 $\times$ 1200 s to 1--2 $\times$ 1800 s and shorter exposures were taken to enable measurement of bright lines. Data were reduced using the 2dfdr pipeline provided by the AAO, however the splicing of blue and red arms was performed separately. 

AAOmega field centres were largely selected to maximise target density, but some centres were modified to include binaries discovered during the course of this project. For 10 of the 12 fields we carefully inspected full resolution SHS and SuperCOSMOS Sky Survey (SSS, Hambly et al. 2001) data using the techniques described by Miszalski et al. (2008a) to select fibre positions for known, MASH and MASH-II PNe. Many uncatalogued objects exhibiting H$\alpha$ emission were assigned spare fibres and this eventuated in $\sim$40 new very low surface brightness MASH-II PNe. These discoveries contributed to our master catalogue (Sec. \ref{sec:catalogue}) and some were covered by OGLE-III. This approach also ensured many PN-mimics were observed (Sec. \ref{sec:mimic}). In most fields a few hundred sky targets were included to assist subtraction of the variable sky background. Prior to observation fibres were assigned to selected positions using the simulated annealing field configuration algorithm (Miszalski et al. 2006). The relatively low target densities per field ($\sim$20--70 PNe) enabled $\sim$200 sky targets and uncatalogued H$\alpha$ emitters to be allocated, although the latter were given lower priority than catalogued PNe.

Additional spectroscopic observations were conducted by AA and BM using the VLT FLAMES multi-object spectroscopy facility (Pasquini et al. 2002) during the ESO visitor mode program 0.79.D-0764(A) on 9--12 June 2007. The 25 arcmin diameter field of view and inclement weather limited the observed sample size. We used the deployable miniature integral-field units (mini-IFUs) that feed 20 optical fibres to sample a 2 $\times$ 3 arcsec$^2$ field of view for each PN within a field. In this work we only make use of one field ($\ell=4$, $b=-3$) that happened to contain three PN-mimics (Sec. \ref{sec:mimic}). Here we only make use of the 30 and 60 minute exposures made with the GIRAFFE LR02 and LR03 filters, respectively, that gave a wavelength coverage of 3948--5074 \AA. Data were reduced using the IRAF task \textsc{dofibers} and relative flux calibration was achieved using the spectrophotometric standard star Feige 66.

In the absence of AAOmega or FLAMES observations we draw upon the extensive spectroscopic resource of the MASH project (Parker et al. 2006; Miszalski et al. 2008a). The predominantly \emph{confirmatory} MASH spectra are not as deep, but are nonetheless useful for bright emission line ratios and in some cases detecting PN-mimics. Of most relevance to this work is the 6--15 May 2008 MASH observing run on the Australian National University 2.3-m telescope. While the focus was to measure radial velocities of a few hundred PNe towards the Bulge via short exposures (Miszalski et al. in prep), a number of the binaries discovered during the course of this project were given special attention. We used the 1200B and 1200R gratings with a slit width of 2\arcsec\, to give wavelength coverage windows of 4030--5050 \AA\, and 6245--7250 \AA\, at a resolution of $\sim$1.6 \AA\, (FWHM). 
Data reduction was performed using IRAF with the assistance of the \textsc{PNDR} package.\footnote{http://www.aao.gov.au/local/www/brent/pndr} 

\section{Search Method}
\label{sec:search}
Our search method involves a number of stages designed to find variable CSPN while incrementally refining the sample of PNe considered in later binary fraction calculations. 
After constructing a catalogue of PNe covered by OGLE-III (Section \ref{sec:catalogue}), identifications of genuine or candidate CSPN are attempted using available near-infrared and optical images (Section \ref{sec:cspnid}). We then discuss the influence of nebular contamination (Section \ref{sec:nebular}) and general systematic effects (Section \ref{sec:systematic}) on time series photometry extracted for all identifications. Combining photometry with available spectroscopy we proceed to remove non-PN contaminants (Section \ref{sec:mimic}), before searching for both periodic (Section \ref{sec:idpvar}) and aperiodic (Section \ref{sec:idavar}) variables indicative of binarity.

\subsection{Planetary Nebulae Catalogue Construction}
\label{sec:catalogue}
Full exploitation of OGLE-III for this work requires a large sample of PNe towards the Galactic Bulge that must be compiled from various extant catalogues. We have therefore carefully constructed a comprehensive PNe sample for $|\ell|\le13$ and $|b|\le7$ to cover all OGLE-III fields with coordinates of each PN verified using the SHS. The Macquarie/AAO/Strasbourg H$\alpha$ PNe catalogues MASH-I (Parker et al. 2006) and MASH-II (Miszalski et al. 2008a) form the foundation of the catalogue with $\sim$380 and $\sim$100 PNe located in the region, respectively. An additional $\sim$450 PNe were then added from a large number of literature sources (e.g. Acker et al. 1992, 1996; Kohoutek 1994, 2002; Zanin et al. 1997; Beaulieu, Dopita \& Freeman 1999; Cappellaro et al. 2001; Jacoby \& Van de Steene 2004; Boumis et al. 2003, 2006; G\'orny 2006). As a further check we made use of updated coordinates provided by Kerber et al. (2003) and SIMBAD to ensure no objects were missed. 

Figure \ref{fig:survey} depicts the most complete catalogue to date of Bulge PNe overlaid on OGLE-III fields. The OGLE field finder\footnote{http://ogle.astrouw.edu.pl/radec2field.html} reported that 297 out of the $\sim$930 PNe were covered by OGLE-III fields with at least one observation. These numbers do not include non-PN contaminants (Sec. \ref{sec:mimic}). The catalogue exhibits considerable variety from extremely low surface brightness MASH PNe to very high surface brightness non-MASH PNe such as NGC 6565.

\subsection{Central Star Identification}
\label{sec:cspnid}
Table \ref{tab:super} presents our sample of 297 PNe for which we have carefully examined all available images to attempt CSPN identification. 
Central to our search was the inspection of $2 \times 2$ arcmin$^2$ OGLE-III $I$-band images. The location and extent of each PN was established using an optical colour-composite image derived from the SHS and SSS with red, green and blue channels constructed from H$\alpha$, Short Red and $B_J$ images, respectively (e.g. Miszalski et al. 2008a). Visual comparisons between the optical and $I$-band images were made using a near infra-red (NIR) colour-composite image as a reference point with red, green and blue channels constructed from the 2MASS $K_s$, $H$ and $J$ atlas images, respectively (Skrutskie et al. 2006). OGLE-III $V$-band images and magnitudes were unavailable given the provisional calibration of Bulge data (Udalski et al. 2008), but for four PNe we have inspected their $V$-band images (effectively shallow [OIII] images).

Each entry in Tab. \ref{tab:super} is remarked according to the type and quality of each identification (parenthesised and quoted labels discussed below). A `true' (T) classification is based on either the presence of a single star at the geometric centre of the nebula or blue colours. Without the $V$-band data we could not search all PNe for blue CSPN with intrinsic colour $(V-I)_0=-0.40$ (Ciardullo et al. 1999). Instead we have constructed SSS $B_J-R_F$ images that are sensitive in some cases to faint blue (B) CSPN (Parker et al. 2006). The 2MASS colour-composite is also useful in which bright CSPN of typically non-MASH PNe appear violet. Some blue CSPN were found differently, e.g. ESO Imaging Survey data (Nonino et al. 1999) were used for PHR 1753$-$3443, and OGLE-III $V$-band images were used for H 2-29 and K 6-34. There are 92 PNe with `true' CSPN identifications. 

In more ambiguous cases where a number of CSPN candidates are present we can only assign `likely' (L) or `possible' (P) based on available data. For a PN marked as `likely' a number of CSPN candidates can be found near the geometric centre of the PN. If `possible', then these candidates may not be near the geometric centre or the centre may be poorly defined. In both cases time series photometry is extracted and examined for all candidates bounded by the H$\alpha$ emission of each PN. In practice distinction between `likely' and `possible' objects is rather difficult as it is complicated by extinction and crowding. Unfortunately, the poor resolution of the often-saturated SSS data means few meaningful colour selections can be made resulting in the number of `true' CSPN identifications being a lower limit until $V$-band data become available. However, in some cases comparison of on- and off-band images from Ruffle et al. (2004) helped to better establish the geometric centre enough to select one CSPN as `true'. There are 102 and 48 PNe with `likely' and `possible' CSPN identifications.

A number of different scenarios occur when no CSPN identification is possible. When the $I$-band image is available and absolutely no CSPN candidates were detected these are marked as `NONE' (2 PNe). A large group of 34 `NULL' PNe include very bright non-MASH PNe avoided by the OGLE-III pipeline, PNe that fell just outside the requested $I$-band image or near a bad CCD column, and PNe for which no time series photometry could be extracted by the pipeline (Sec. \ref{sec:systematic}). Nebular detections are indicated by `NEB' for 10 PNe \emph{only} when there is a nebular rim or other faint detection not coincident with a CSPN (e.g. NGC 6565, M 1-29).  No label is given for all nebular detections but those indicated by `NEB' are in the minority (see Section \ref{sec:nebular}). Suspiciously spurious lightcurves caused by systematic effects are indicated by `S' for 14 PNe (see Sec. \ref{sec:systematic}). Periodic variables are indicated by `BIN' (binary)  and aperiodic variables by `BIN?' (suspected binary).

\subsection{Nebular contamination}
\label{sec:nebular}
\begin{figure}
   \begin{center}
      \includegraphics[scale=1.0,angle=270]{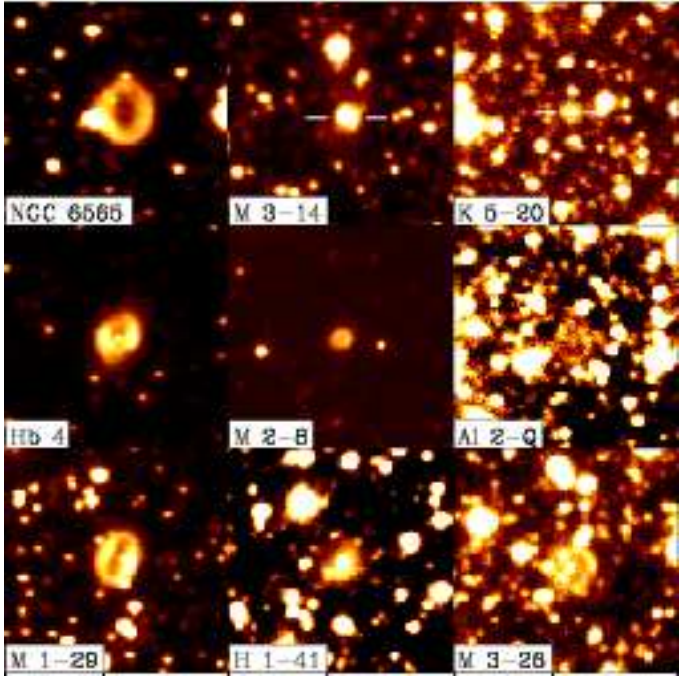}
   \end{center}
   \caption{A selection of OGLE-III $I$-band images of PNe with nebular detections. All images are 30 $\times$ 30 arcsec$^2$ with North to top and East to left. 
   }
   \label{fig:sample}
\end{figure}

Nebular detections occur for $\sim$20\% of the PNe in our sample to varying degrees due to the strong nebular [SIII] emission lines passed by the broad OGLE-III $I$-band filter (see Sec. \ref{sec:ogle}). 
Figure \ref{fig:sample} shows a sample of nebular detection levels in our sample. 
Notably absent from Fig. \ref{fig:sample} are MASH and MASH-II PNe of which almost none have nebular detections because they are intrinsically much fainter than non-MASH PNe. 

In many nebular detections the nebulae appear faint or diffuse (e.g. K 5-20) and prove to be no obstacle to the \emph{stellar-optimised} OGLE-III pipeline. Problems can occur however when the nebula becomes bright and large enough to appear non-stellar. Fully resolved nebulae such as M 1-29 and NGC 6565 cannot be treated in any reasonable fashion (remarked `NEB' in Tab. \ref{tab:super}). These nebulae may also confuse the pipeline even if a CSPN is clearly visible (e.g. Hb 4). 
Between these two extremes are objects such as M 2-8 and M 3-14 where a bright and just resolved nebular core is present. The non-stellar appearance of these objects means extraction of time series photometry by the pipeline is not guaranteed. If photometry is obtained then we assume the nebula is obscuring an underlying CSPN and that we are sensitive to any variations present. This is a reasonable assumption given the brightness of the objects and indeed we have found one binary belonging to this description (H 1-33, see later). These objects may however be more prone to systematic effects as discussed in the next section. 

\subsection{Systematic effects}
\label{sec:systematic}
\begin{figure}
   \begin{center}
      \includegraphics[scale=0.6,angle=270]{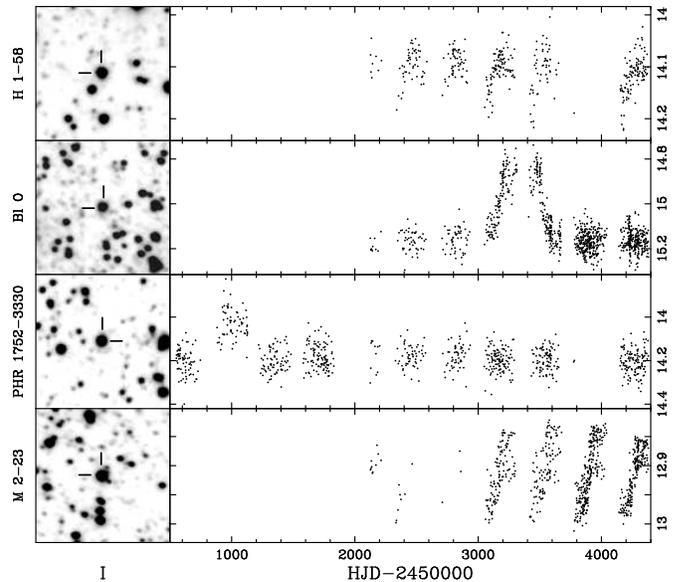}
   \end{center}
   \caption{Examples of suspected spurious time series photometry. H 1-58 and M 2-23 show distinctly regular one year periods. Bl O and PHR 1752-3330 (prefixed by OGLE-II data) show unusually long sustained `outbursts'. 
   }
   \label{fig:spurious}
\end{figure}

The time series photometry of $\sim$14 PNe appear to be dominated by systematic effects producing spurious periodicities of one year or non-repeating `outbursts' (Fig. \ref{fig:spurious}). 
While systematic effects in OGLE-III are generally very small permitting transiting planet discovery, they can become prominent and difficult to explain in extreme non-stellar objects such as PNe that show strong nebular contamination. In the most extreme PNe, i.e. many of those marked as `NULL', no time series photometry could be secured by the \emph{stellar-optimised} pipeline (Wo\'zniak 2000). In practice only a few PNe may be rejected based on strong nebular contamination alone. The majority of `NULL' PNe can rather be attributed to a very bright object being avoided by the pipeline or a bad CCD column across the PN in a small number of cases. Those that do pass the pipeline with significant nebular contamination may be more prone to systematic effects.
Wo\'zniak (2002) estimated 10\% of OGLE-II Bulge variables are spurious having been caused by problems undetected at the pipeline level (Wo\'zniak 2000). Some documented sources of systematic errors include colour effects (Udalski et al. 2002b) and `mirrored' variability from bright variables up to 100 pixels away (Mizerski \& Bejger 2002).

An exact cause of the suspect variability in Fig. \ref{fig:spurious} is unclear based on the available data. The most likely cuplrit is differential atmospheric refraction (DAR) exacerbated by strong nebular contamination. DAR is the shift in centroid position of a star dependent on the colour of the star and the airmass of the observation (Alcock et al. 1999a, 1999b).
When the nebula point-spread function is considerably larger than regular stars we suspect a small shift in the real centroid, compared to the fixed reference image, to cause quite large light variations. Indeed, the `inverted-U' shape present in H 1-58 and M 2-23 appears related to airmass changes as the position of the Bulge in the sky changes during each observing season. Similar features are mentioned by Alcock et al. (1999a) as `slow seasonal rolls' with one year periods.

Photometric outbursts may be related to transient mass-loss events as in Lo 4 (Bond 2008), but we discount this explanation for Bl O and PHR 1752$-$3330 given the 1--2 year non-repeating duration is considerably longer than Lo 4 which is active $\sim$8\% of the time (Bond 2008). This is supported by inspection of different time series obtained at opposing edges of the fully resolved nebula of Bl O (Ruffle et al. 2004) where we found the two `outburst' peaks to suspiciously halve in magnitude alternately on each edge. 

Another source of systematic error pertains to those light curves that show considerably large scatter $\sigma$ despite relatively small error bars of individual data points. This may be attributed to non-photometric or cloudy nights rather than large intrinsic variability (Udalski et al. 2008). We cleaned our sample by first selecting objects with large $\sigma$ as a function of average magnitude (amongst all identified CSPN), and then checking whether the variability persisted when compared to stars surrounding each object. This process showed the vast majority of these cases were not caused by intrinsic variability (see also Sec. \ref{sec:idavar}).

At this stage it suffices to identify any time series photometry with suspicious one year periodicity or outbursts. These have been remarked with an `S' or `S?' in Tab. \ref{tab:super}. We refrain from marking PHR 1752$-$3330 as such given the outburst occurs for OGLE-II data only. Further investigation into systematic effects may be possible once $V$-band images and magnitudes become available to investigate the role of colour in DAR.

\subsection{Removing PN Mimics}
\label{sec:mimic}
A significant number of PN mimics are present in existing PN catalogues (for an excellent review see Frew 2008; see also Parker et al. 2006, Miszalski et al. 2008a). Mimics are often mistaken for PN based solely on their H$\alpha$ emission without any deep follow-up spectroscopy to confirm their nature. 
We cannot overemphasise the importance of removing mimics to both reduce the possibility of identifying non-PN binaries and  to obtain an accurate estimate of the binary fraction. We were able to identify and remove mimics based on 2MASS colours and deep FLAMES and AAOmega spectroscopy of identified H$\alpha$ emitters in many fields overlapping with highly sampled OGLE-III fields (Fig. \ref{fig:survey}).
Table \ref{tab:mimic} contains the mimics identified in OGLE-II and OGLE-III fields that are discussed in the following sections.

\begin{table*}
   \centering
   \begin{tabular}{llllrrrrrll}
      \hline\hline
      PN G& Name & RA (J2000) & DEC (J2000) & $R1$ & $R2$ & $J-H$ & $H-K_s$ & $K_s$ & Remarks & Ref.\\ 
      \hline
  006.1$+$04.1 & - & 17 44 10.6 & $-$21 29 21 & - & -& 1.24 & 1.91 & 11.10               & Sy?,H,P     & (f)     \\              
  356.1$-$02.1 & PHR 1744$-$3319 & 17 44 51.8 & $-$33 19 30 & - & - & - & - & - & Sy? & (a) \\  
  355.5$-$02.8 & MPA 1746$-$3412& 17 46 18.5 & $-$34 12 37 & - & - & - & - & -                       & Sy,A,OVI      & (b)     \\            
  355.0$-$03.3 & PPA 1746$-$3454  & 17 46 51.4 & $-$34 54 05 & 0.76 & 0.94 & 1.96 & 2.04 & 11.45    & Sy,A,H,OII    & (a)     \\           
  359.5$-$01.3 & JaSt 68 & 17 49 50.9 & $-$30 03 10 & - & - & - & - & - & U,A & (c)\\  
  000.0$-$01.1  & JaSt 2-6 & 17 50 01.0 & $-$29 33 25 & -& -& 2.76 & 1.97 & 6.16                    & Sy?,N,OII,P & (c),(d) \\ 
  356.4$-$03.5 & PHR 1751$-$3349 & 17 51 15.0 & $-$33 49 11 & - & - & 2.30 & 1.72 & 9.99 & Sy?,A,N & (b) \\ 
  358.0$-$02.7 & Al 2-O & 17 51 45.4 & $-$32 03 04 & - & 9.3 & 0.98 & 0.89 & 10.84 & Sy?,P & (i)\\
  000.2$-$01.4 & JaSt 79         & 17 51 53.5 & $-$29 30 53 & 1.4 & 11.2 & 1.65 & 1.14 & 7.11                & Sy,A,F,M       & (c)     \\       
  354.8$-$04.6 & PPA 1752$-$3542 & 17 52 05.9 & $-$35 42 06 & 1.8 & 2.7 &- &- & -                    & Sy,P        & (a)     \\     
  000.6$-$02.3 & H 2-32          & 17 56 24.2 & $-$29 38 07 & - & 0.1 & 0.38 & 0.86 & 12.57 & Sy?,A,H,P & (j) \\
  356.2$-$05.1 & K 5-37          & 17 57 15.7 & $-$34 47 34 & - & - & 0.09 & 1.82 & 11.86 & Sy?         & (l) \\
  001.0$-$02.6 & Sa 3-104        & 17 58 25.9 & $-$29 20 48 & 0.5 & 1.6 & 0.72 & 0.80 & 11.81 &  Sy,A,OII & (h)\\ 
  003.4$-$01.9 & MPA 1801$-$2655   & 18 01 06.3 & $-$26 55 59 & 0.7 & 1.2 & 1.19 & 0.62 & 8.02    & Sy,A,H,M    & (b)     \\          
  356.9$-$05.8 & M 2-24        & 18 02 02.9 & $-$34 27 47 & 1.4 & 3.7 & 1.78 & 1.83 & 9.31        & Sy,N      & (e)     \\              
  002.9$-$02.7 & PHR 1803$-$2746 & 18 03 05.1 & $-$27 46 44 & 3.0 & 8.1 &- &- & -                    & Sy,A        & (a)     \\            
  003.0$-$02.8 & PHR 1803$-$2748 & 18 03 31.2 & $-$27 48 27 & 0.2 & 0.5  & 1.05 & 2.16 & 10.56     & Sy,A,H,N,OVI  & (b)     \\            
  001.2$-$03.9 & ShWi 5& 18 03 53.8 & $-$29 51 22 & 1.5 & 9.4  & 0.60  & 0.60 & 11.90              & Sy,A,F      & (g)     \\                      
  001.7$-$03.6 & MPA 1804$-$2918     & 18 04 04.9 & $-$29 18 46 & 0.1 & 0.2 & 0.88 & 0.34  & 7.78 & Sy,A,H,M,OII   & (b)     \\   
  001.7$-$03.8 & ShWi 7          & 18 05 05.5 & $-$29 20 15 & 0.2 & 4.2 & 0.67 & -0.01 & 13.02 & Sy?,A,N,OII & (g)\\
  003.9$-$02.8 & PHR 1805$-$2659 & 18 05 43.5 & $-$26 59 46 & -& -& 0.50 & 0.42 & 10.96               & F$*$,A,M  & (a)     \\        
  004.0$-$03.0 & M 2-29          & 18 06 40.8 & $-$26 54 56 & 1.0 & 4.0 & 0.41 & 1.17 & 12.32 & Sy?,H,N,OII,V  & (k) \\ 
  004.1$-$03.0 & PHR 1806$-$2652 & 18 06 56.0 & $-$26 52 54     & 0.1 & 4.1 & 0.95 & 0.64 & 12.90  & Sy,F,M,N,OII,OVI?,V& (a)     \\            
  359.7$-$05.5 & PPA 1807$-$3158 & 18 07 19.7 & $-$31 58 09     &- &- & 0.41 & 1.69 & 12.31           & Sy?,P      & (a)     \\            
  004.1$-$03.3 & PPA 1808$-$2700 & 18 08 01.4 & $-$27 00 16 & 0.5 & 3.8 & 1.03 & 0.30 & 9.18   & Sy,M,N,OII,V  & (a)     \\        
  003.6$-$04.0 & PHR 1809$-$2745 & 18 09 51.8 & $-$27 45 54     & -&- & 1.15 & 1.75 & 9.32            & Sy?,OVI,P     & (a)     \\            
   \hline
   \end{tabular}
   \begin{flushleft}
      \emph{Remarks:}\\  
      Sy(?): Symbiotic star (suspected); F$*$: Flare star; U: Unlikely PN; A: AAOmega observations \\
      F: [FeVII] $\lambda\lambda$5721, 6087 present; H: Broad H$\alpha$; M: Late-type features; N: Resolved Nebula\\
      OII: OGLE-II coverage; OVI: OVI $\lambda$6830 present; P: Other MASH observations; V: FLAMES observations\\
      \emph{References:}\\
      (a) Parker et al. (2006), (b) This work, (c) Jacoby \& Van de Steene (2004), (d) Matsunaga, Fukushi \& Nakada (2005)\\
      (e) Zhang \& Liu (2003), (f) G\'orny (2006), (g) Shaw \& Wirth (1985), (h) Sanduleak (1976), (i) Allen (1979)\\
      (j) Haro (1952), (k) Hajduk et al. (2008), (l) Kohoutek (2002)\\
   \end{flushleft}
   \caption{PN mimics removed from our PN sample. Columns $R1$ and $R2$ are the emission line ratios [OIII] $\lambda$4363/H$\gamma$ and [OIII] $\lambda$5007/H$\beta$, respectively (Guti\'errez-Moreno, Moreno \& Cort\'es 1995). NIR magnitudes are from 2MASS (Skrutskie et al. 2006). 
   }
   \label{tab:mimic}
\end{table*}

\subsubsection{Symbiotic Stars}
\label{sec:sy}
In a photometric search for binary CSPN the main source of contaminants are symbiotic stars. 
Symbiotic stars are binary systems typically composed of a late-type giant or Mira transferring material to a hot white dwarf companion via a stellar wind (Miko{\l}ajewska 2003, 2007) and they exhibit large and varied photometric variability (Miko{\l}ajewska 2001; Gromadzki, Miko{\l}ajewska \& Borawska et al. 2007).  Many are associated with extended nebulae (Corradi et al. 1999, 1995) that are easily mistaken for PN but are not the ejecta of the hot component as in PNe (Corradi 2003). Their PN-like nebulae can even exhibit peculiar variability (Doyle et al. 2000; Shaw et al. 2007).

NIR colours can be very useful in identifying symbiotics (Schmeja \& Kimeswenger 2001; Phillips 2007; Corradi et al. 2008), but ultimately deep spectroscopy is required to identify typically faint features needed for definitive classification. 
These features include Raman-scattered OVI emission lines $\lambda\lambda$ 6830, 7082 unique to symbiotics (Schmid 1989), [OIII] $\lambda$4363 useful in separating PN from symbiotics in the $R1=$ [OIII] $\lambda$4363/H$\gamma$, $R2=$ [OIII] $\lambda$5007/H$\beta$ plane (Guti\'errez-Moreno, Moreno \& Cort\'es 1995) and stellar features typical of the late-type companion (Munari \& Zwitter 2002; Belczy\'nski et al. 2000). Indicative values of $R1\ga0.6$ and $R2\la10$ are a reasonable signature of a symbiotic star but additional criteria should be satisfied for a definitive classification.

\begin{figure}
   \begin{center}
      \includegraphics[scale=0.5]{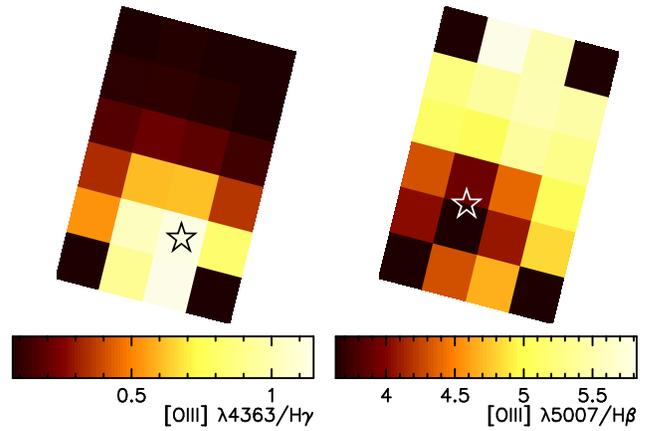}
   \end{center}
   \caption{Spatial distribution of M 2-29 emission line ratios $R1$ (left) and $R2$ (right) suggestive of a symbiotic star. A star marks the approximate central star position that shifts in each exposure because of differential atmospheric diffraction. The FLAMES mini-IFU field of view is 2 $\times$ 3 arcsec$^2$ with North to top and East to left.}
   \label{fig:m2-29}
\end{figure}

\begin{figure}
   \begin{center}
      \includegraphics[angle=270,scale=0.6]{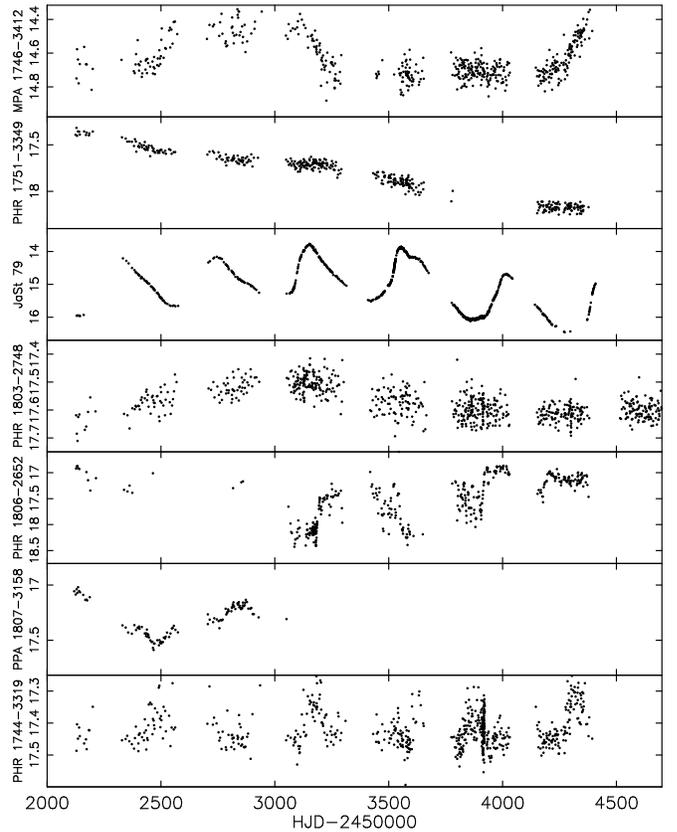}
   \end{center}
   \caption{A selection of OGLE-III time series photometry for confirmed and strongly suspected symbiotic stars removed from our PNe sample. }
   \label{fig:sylc}
\end{figure}

During the catalogue construction phase (Sec \ref{sec:catalogue}) well-known symbiotics were removed using SIMBAD and literature sources (Belczy\'nski et al. 2000; Phillips 2007; Schmeja \& Kimeswenger 2001). Application of the above criteria to the penultimate catalogue produced the cleaned final sample of 297 PNe (Tab. \ref{tab:super}). We found the depth and large wavelength coverage of our AAOmega data especially helpful in identifying symbiotics. Almost all objects in Tab. \ref{tab:mimic} are newly confirmed (Sy) or strongly suspected (Sy?) symbiotic stars and many were previously classified as PNe. 
Those presented here for the first time, with this work as the reference, were identified during the construction of the MASH (PHR objects) or MASH-II (MPA objects) catalogues. M 2-29 has been classified as a PN harbouring a triple system (Hajduk et al. 2008), but we suspect a symbiotic star after our spatially resolved FLAMES spectra confirmed high $R1$ and low $R2$ values near the central star (Fig. \ref{fig:m2-29}) that were first reported by Torres-Peimbert et al. (1997). However, there may be another explanation for these line ratios and our tentative classification will be refined in a future paper (Miszalski et al., in prep). Figure \ref{fig:sylc} presents a sample of the extreme variability seen in these new symbiotics typified by irregular and long term variations. Outside OGLE fields we have reclassified more PNe as symbiotics that will be presented elsewhere along with more detailed analysis of the symbiotics in Tab. \ref{tab:mimic}.

\subsubsection{PHR 1805$-$2659: An M-dwarf flare star}
\begin{figure}
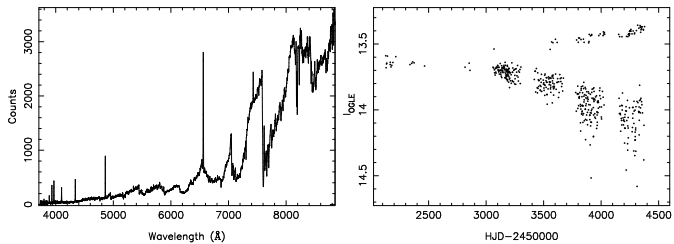

   \begin{center}
      \includegraphics[angle=270,scale=0.19]{PHR1805_2659_spectrum.ps}
      \includegraphics[angle=270,scale=0.19]{PHR1805_2659_lc.ps}
   \end{center}
   \caption{AAOmega spectrum (left) and OGLE-III time series photometry (right) of the newly identified M-dwarf flare star PHR 1805$-$2659.}
   \label{fig:flare}
\end{figure}

We reclassify the `possible' MASH PN PHR 1805$-$2659 as an M-dwarf flare star (e.g. Doyle et al. 1990). Figure \ref{fig:flare} shows the 1800 s AAOmega spectrum taken on 29 May 2008 and OGLE-III time series photometry of the object. The M-dwarf spectrum exhibits strong Hydrogen Balmer and Ca H and K emission lines consistent with intense chromospheric activity. The activity is reflected in the OGLE-III photometry as an increasing sequence of flares separate to a declining quiescent sequence.

\subsection{Periodic variable determination}
\label{sec:idpvar}
The lomb-scargle technique was used as our main method for finding periodic variables (Lomb 1976; Scargle 1982; Press et al. 1992). 
A \textsc{matlab} implementation of the \textsc{period} function (Press et al. 1992) was acquired and modified to use \textsc{pgplot} to generate periodograms covering a long period window (5--100 d) and a short period window (0.1--5 d) under \textsc{gnu octave}.\footnote{http://www.gnu.org/software/octave} The periodograms of all stars within the radius of each PN were visually inspected for peaks with a significance level of $\sim$1\% or less. There is a small chance that an unrelated variable star could be found within a PN (e.g. Sh 2-71, Frew \& Parker 2009), however it is relatively uncommon to find more than one variable star per PN. We emphasise that if there are any of these objects in our search, their assignment as likely or possible CSPN identifications means they are treated appropriately in our estimate of the binary fraction (see later). We checked the results using both \textsc{period04} (Lenz \& Breger 2004) and less often the IRAF task \textsc{pdm} (Stellingwerf 1978) with excellent agreement found amongst all three methods.
Periods on the order of a hundred days were better searched for visually as discussed in the following section.

\subsection{Aperiodic variable determination}
\label{sec:idavar}
We investigated outliers in the magnitude dispersion $\sigma$ versus average magnitude plane for all `true' CSPN with extracted time series photometry.
The vast majority of outliers were found to be spurious after comparison with surrounding stars. A few objects showed real variation but most were symbiotics. Indeed, unduly large variations can be used as a good tool in revealing symbiotics in conjunction with other data. All photometry was also inspected visually to identify any unusual variations with the time axis `stretched-out' to better resolve the `bunched-up' Bulge seasons.

\section{Results and Discussion}
\label{sec:results}
\subsection{New Binary Central Stars}
\label{sec:binaries}
Table \ref{tab:periodic} presents 22 binary CSPN resulting from our searches with 19 of these presented here for the first time. We include M 3-16, H 2-29 and M 2-19 from Miszalski et al. (2008b). All discoveries are from OGLE-III except for JaSt 66 which is an OGLE-II discovery. A `C' is placed in the CSPN column of Tab. \ref{tab:periodic} where spectroscopic confirmation of the CSPN has been achieved, otherwise the TLP identification is given (see Sec. \ref{sec:cspnid}). 
We emphasise that the `T' and `L' binaries still require spectroscopic confirmation.
The `Type' column uses nomenclature adopted from DM08 to describe the cause of variability as either irradiation effects (I), ellipsoidal variation (El) or eclipses (Ec). The epoch of primary minimum $E_0$, the period $P$ and peak-peak amplitude $A$ (only for irradiated binaries) were calculated using \textsc{period04} (Lenz \& Breger 2004). Periods for the seven eclipsing systems were selected to give two minima per cycle as usual. 
When two different minima per cycle were found we put this down to ellipsoidal variation. 
Otherwise we have assumed irradiation effects are the cause of the variation and select periods that show only one minimum per cycle. Future RV observations may possibly double the period of a few irradiated variables to become ellipsoidal variables, but for now we settle on the periods given in Tab. \ref{tab:periodic}. The average OGLE-III $I$-band magnitudes ($\bar{m}_{I}$) are accurate to 0.1--0.2 mag (Udalski et al. 2008). More uncertain magnitudes are marked with `:' including K 6-34 where we give the average magnitude of the two close stars adding $\sim$0.2 mag of uncertainty (see below) and H 1-33 which is affected by moderate nebular contamination.

\begin{table*}
   \centering
   \begin{tabular}{lllllllllll}
      \hline\hline
      Field & PN G & Name & CSPN & Type& $E_0$ & $P$ (days) & $\bar{m}_I$ & A (mmag) \\
      \hline
      BLG129 &  355.6$-$02.3 & PHR 1744$-$3355&T  & I        & 3575.20372 & 8.233928 & 16.5 & 97 \\
      BLG342 &  005.0$+$03.0 & Pe 1-9         &T  & Ec,I?    & 2934.58226 & 0.139858 & 17.6 & - \\
      BLG121 &  355.3$-$03.2 & PPA 1747$-$3435&T  & I    & 4235.85512 & 0.224709 & 18.0 & 119     \\
      BLG130 &  355.7$-$03.0 & H 1-33         &T  & I        & 3867.94848 & 1.128491 & 16.9:    & 245  \\
      BLG117 &  354.5$-$03.9 & Sab 41         &C  & I        & 2711.74765 & 0.297155 & 16.5 & 1698 \\
      BUL\_SC44&  359.5$-$01.2 & JaSt 66      &T  & El   & 1772.62919 & 0.275557 & 17.8 & -    \\
      BLG195 &  000.6$-$01.3 & Bl 3-15        &T  & El   & 3310.44171 & 0.270218 & 18.2 & - \\
      BLG180 &  359.1$-$02.3 & M 3-16         &C  & Ec,El & 3850.91841 & 0.573648 & 15.9 & -    \\
      BLG155 &  357.6$-$03.3 & H 2-29         &T  & Ec,I?    & 3619.52078 & 0.244110 & 18.1 & -\\ 
      BLG101 &  000.2$-$01.9 & M 2-19         &C  & El   & 4224.78224 & 0.670170 & 15.9 & -    \\
      BLG172 &  358.7$-$03.0 & K 6-34         &C  & El       & 3525.59585 & 0.393309 & 16.4: & -    \\
      BLG142 &  357.0$-$04.4 & PHR 1756$-$3342&C  & I        & 4331.41964 & 0.265733 & 18.0 & 551 \\
      BLG214 &  001.8$-$02.0 & PHR 1757$-$2824&T & Ec,I      & 4286.62547 & 0.799209 & 18.1 & 453 \\
      BLG104 &  001.2$-$02.6 & PHR 1759$-$2915&L  & Ec       & 4524.84332 & 1.103664 & 17.5 &  -   \\
      BLG188 &  000.5$-$03.1a& MPA 1759$-$3007&L   & El      & 3930.67937 & 0.503604 & 18.2 & -    \\
      BLG215 &  001.9$-$02.5 & PPA 1759$-$2834&L  & I        & 3099.73716 & 0.305848 & 18.2 & 501  \\
      BLG135 &  357.1$-$05.3 & BMP 1800$-$3408&T  & Ec,I     & 2934.58226 & 0.144777 & 18.1 & 80   \\
      BLG189 &  000.9$-$03.3 & PHR 1801$-$2947&L  & I        & 2132.58866 & 0.315998 & 17.5 & 110  \\
      BLG151 &  357.9$-$05.1 & M 1-34         &L  & Ec      & -          & -        & 16.8 &  -   \\
      BLG233 &  003.1$-$02.1 & PHR 1801$-$2718&L  & I        & 4339.47776 & 0.321972 & 17.3 & 283  \\
      BLG196 &  001.8$-$03.7 & PHR 1804$-$2913&C  & I        & 3578.83476 & 6.659968 & 15.0 & 13\\
      BLG241 &  004.0$-$02.6 & PHR 1804$-$2645&T  & Ec       & 3419.84190 & 0.624505 & 18.5 &  -   \\
      \hline
   \end{tabular}
   \begin{flushleft}
   \end{flushleft}
   \caption{The 22 confirmed, true and likely binary CSPN discovered in this work and Miszalski et al. (2008b). See text for column descriptions. }
   \label{tab:periodic}
\end{table*}

\begin{figure*}
   \begin{center}
      \includegraphics[scale=0.50]{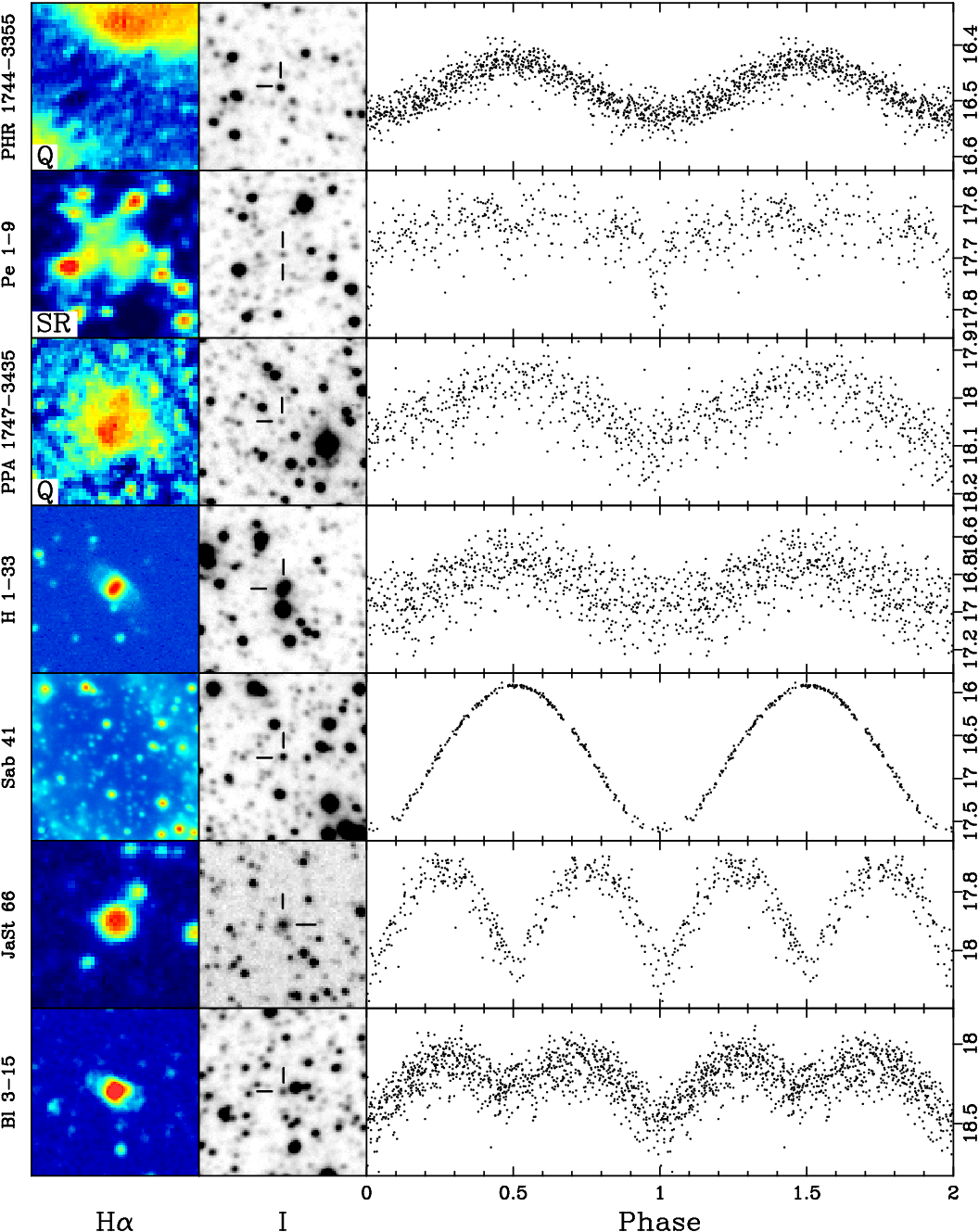}
      \includegraphics[scale=0.50]{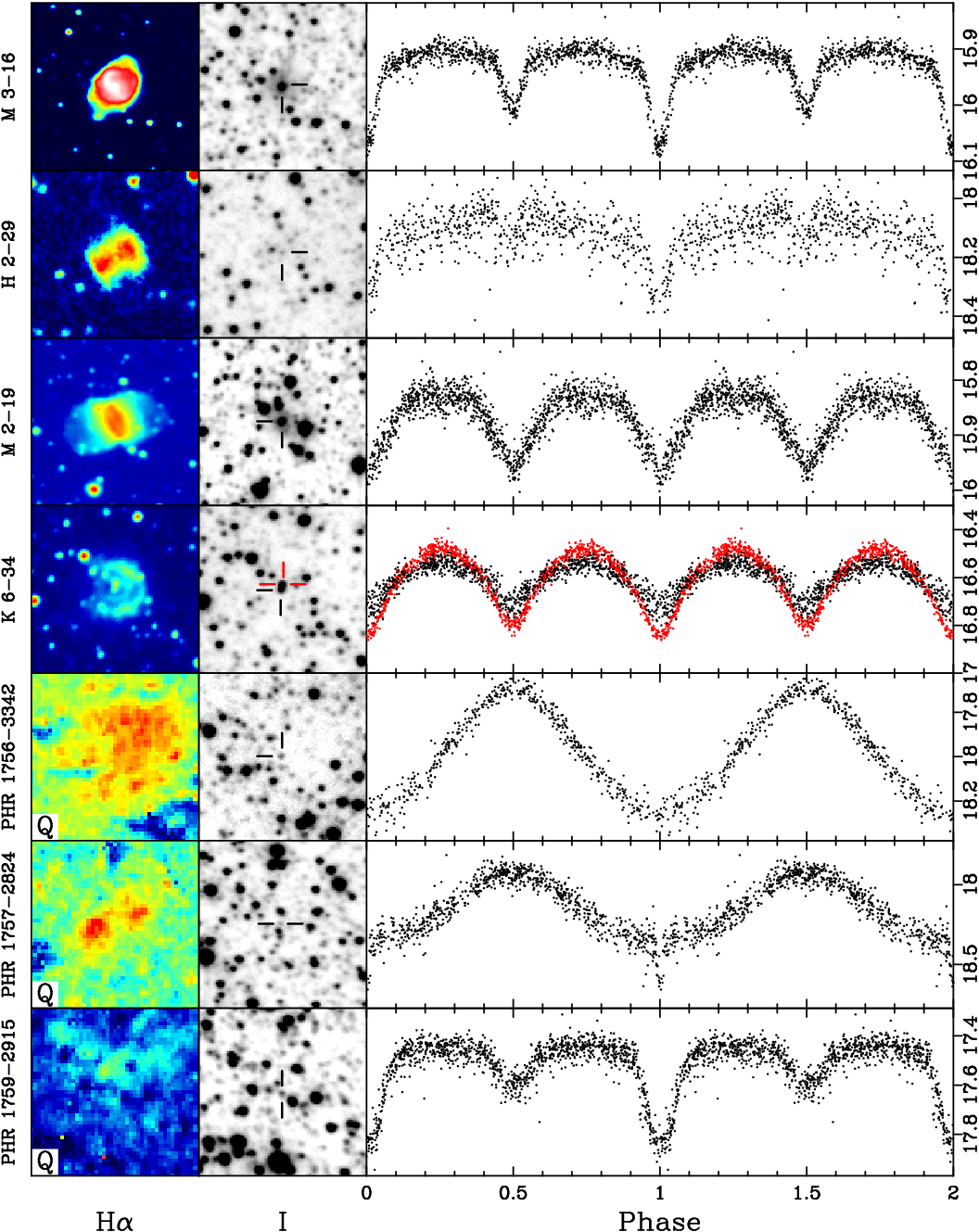}\\
      \includegraphics[scale=0.50]{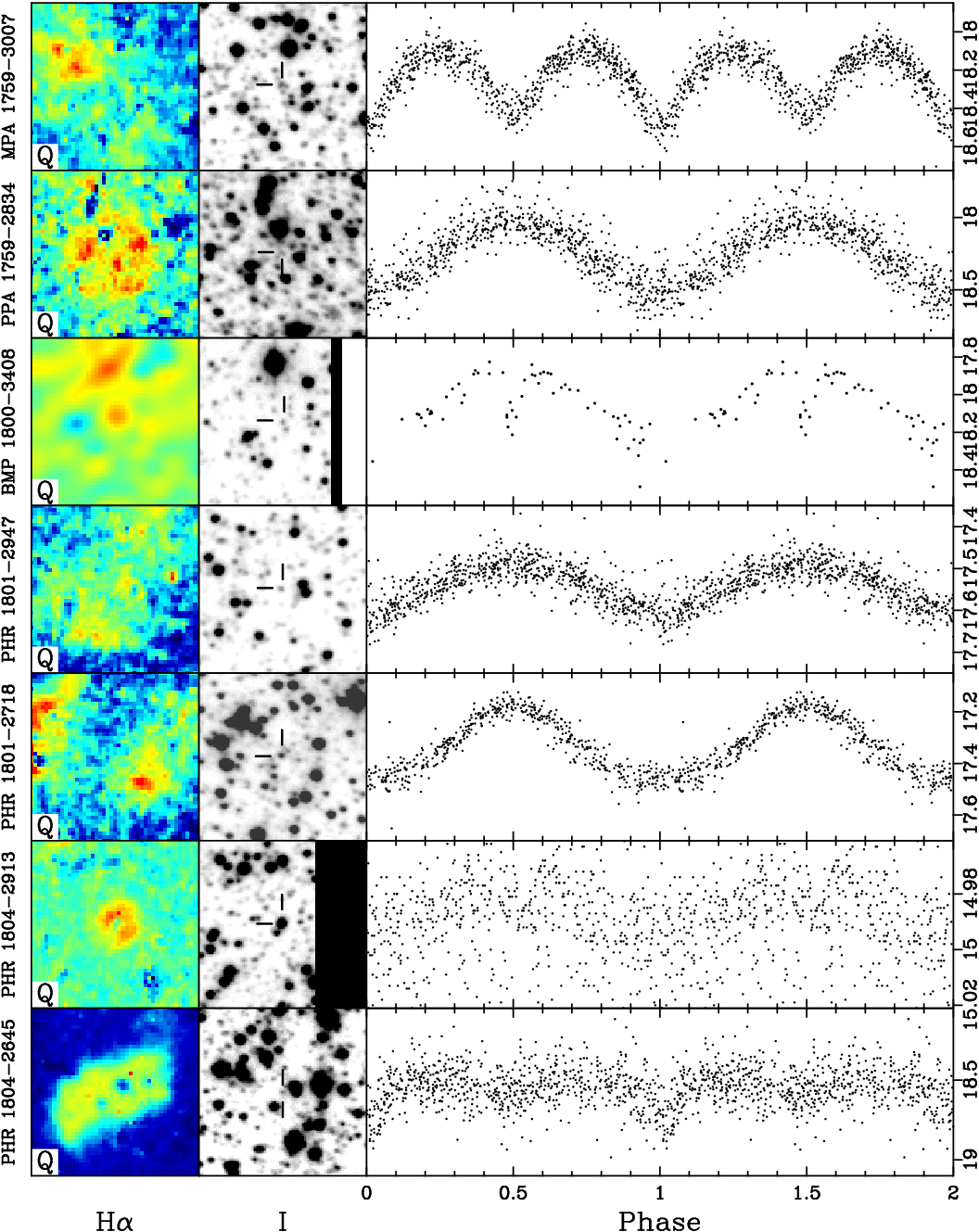}
   \end{center}
   \caption{Images and phased lightcurves of the 21 confirmed, true and likely periodic binary CSPN. 
   Each row per panel depicts left to right: an H$\alpha$ image of the nebula, an OGLE $I$-band image, and a phased OGLE $I$-band lightcurve of the central star (marked on the $I$-band image). Periods and ephemerides are given in Tab. \ref{tab:periodic}. 
   Two indistinguishable lightcurves are given for K 6-34 (see text). 
All images are 30 $\times$ 30 arcsec$^2$ with North to top and East to left.
A `Q' designates an H$\alpha$ quotient image and `SR' an H$\alpha$ off-band image (see text).
   }
   \label{fig:periodic}
\end{figure*}

Figure \ref{fig:periodic} depicts images and phased lightcurves for the 21 periodic binaries from Tab. \ref{tab:periodic}. 
H$\alpha$ images measuring 30 $\times$ 30 arcsec$^2$ are sourced from CCD photometry where available as follows. 
During our ESO visitor mode program 079.D-0764(B) we obtained the 30 s H$\alpha$ exposure of H 1-33 with NTT/EMMI on 21 June 2007. 
Gemini GMOS South (Hook et al. 2004) acquisition images of Sab 41, M 3-16 and M 2-19 were obtained by us using the H$\alpha$ filter on 24, 25 and 31 July 2008 during Program ID GS-2008B-Q-65. Kovacevic \& Parker (private communication) obtained the H$\alpha$ image of K 6-34 using the Mosaic II camera on the CTIO 4-m Blanco telescope on 13 June 2008. The K 6-34 H$\alpha$ image when compared to the OGLE-III $V$- and $I$-band images shows the bluer southern star of a pair separated by 0.9\arcsec\, is the true CSPN. We rule out any physical association with the northern star based on radial velocities, but the close proximity of these stars means the OGLE-III photometry is indistinguishable between the two (illustrated in Fig. \ref{fig:periodic} by the red lightcurve matched to the fainter star). Ruffle et al. (2004) acquired H$\alpha$ images of Bl 3-15 and H 2-29 with NTT/EMMI under ESO program 67.D-0527(A) that we have smoothed and reproduced here. For the remaining 14 PNe we use data from the SHS, mostly as quotient images constructed by dividing H$\alpha$ by the contemporaneous broad-band Short-Red exposure (indicated by `Q' in Fig. \ref{fig:periodic}). For Pe 1-9 we use the Short-Red only because the H$\alpha$ image is saturated. Larger images and a more detailed appraisal of nebular morphologies will be presented in another paper in this series (Miszalski et al., in prep).

\begin{figure}
   \begin{center}
      \includegraphics[scale=0.5]{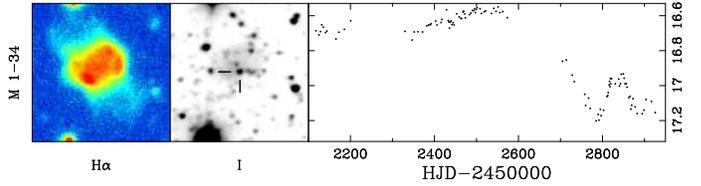}
   \end{center}
   \caption{H$\alpha$ image, OGLE-III $I$-band image and time series photometry of M 1-34.
   Images are 30 $\times$ 30 arcsec$^2$ with North to top and East to left.
   }
   \label{fig:m1-34}
\end{figure}

Figure \ref{fig:m1-34} presents M 1-34 as the only genuine aperiodic variable found and we consider it a likely binary. Other aperiodic candidates were ruled out as either symbiotic stars (Sec. \ref{sec:sy}) or were caused by systematic effects (Sec. \ref{sec:systematic}). The H$\alpha$ image reproduced here was acquired by Ruffle et al. (2004) with NTT/EMMI under ESO program 71.D-0448(A). The star immediately to the NE of our chosen CSPN is not variable, but we cannot rule it out as the CSPN given its slightly more plausible central position. Longslit spectroscopy of both stars is required to clarify the situation. Further photometric monitoring of this object is especially important as the OGLE-III field is no longer observed by the survey.

\subsection{Period Distribution and Selection Effects}
\label{sec:pdist}
Numerous population synthesis calculations have predicted the observed period distribution of close binary central stars (de Kool 1990, 1992; de Kool \& Ritter 1993; Yungelson, Tutukov \& Livio 1993; Han, Podsiadlowski \& Eggleton 1995). 
Although there are many assumptions built into these models, only the initial mass ratio distribution (IMRD) and $\alpha_{CE}$ markedly influence the predicted period distributions. The models therefore offer a valuable tool to directly constrain the IMRD and $\alpha_{CE}$ when compared to the observed period distribution of binary CSPN.
Until now these comparisons have been severely limited by the very small sample size and especially the uncertain observational biases that constituted the sample (DM08). The well-understood observational biases and threefold increase in sample size provided by our OGLE-III discoveries addresses both concerns enabling a fresh and more meaningful comparison to be made.

\begin{figure}
   \begin{center}
      \includegraphics[scale=0.35,angle=270]{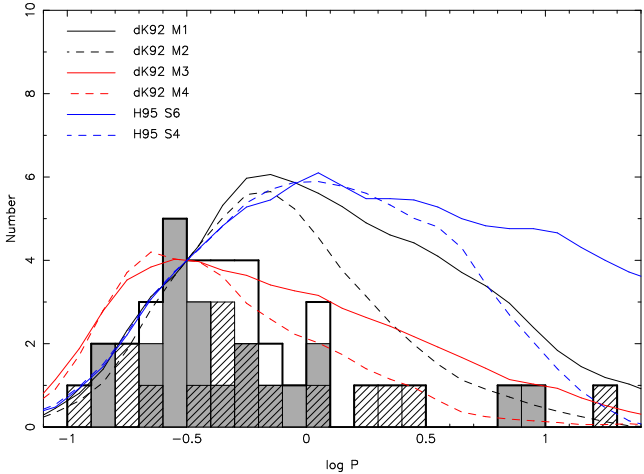}
   \end{center}
   \caption{Orbital period distribution of new binaries (shaded, this work), known binaries (hatched, DM08) and the total sample (thick line). Details of the overlaid models and their normalisation may be found in Tab. \ref{tab:models} and in the text. 
   }
   \label{fig:pdist}
\end{figure}

Figure \ref{fig:pdist} depicts the period distributions of our new sample, the known binaries excluding the uncertain SuWt 2 and Hb 12 (DM08) and the combined distribution. Their appearance is rather sensitive to the bin size for which we have chosen log $P=0.1$. Table \ref{tab:models} outlines the six different models from de Kool (1992) and Han et al. (1995) that have been overlaid on the observed distribution. Han et al. (1995) models make use of an additional $\alpha_{th}$ parameter that has the effect of increasing $\alpha_{CE}$ when $\alpha_{th}=1.0$. Note that the models with lower $\alpha_{CE}$ values (dotted lines) have steeper slopes past the distribution maximum and that the position of the distribution maximum is strongly affected by the IMRD, i.e. dK92 M3 and M4 use a random IMRD, while the other models use a constant IMRD i.e. $\mathrm{d}N\propto\mathrm{d}q$ (de Kool 1992). Other models are available for comparison but these overwhelmingly utilise a constant IMRD and are fairly well represented by some of the models already shown here. 

\begin{table}
   \centering
   \begin{tabular}{lllll}
      \hline\hline
      Model & IMRD & $\alpha_{CE}$ & $\alpha_{th}$ & $\chi^2$ \\
      \hline
      dK92 M1 & $\mathrm{d}N\propto\mathrm{d}q$ & 1.0 & -   & 1.65 \\
      dK92 M2 & $\mathrm{d}N\propto\mathrm{d}q$ & 0.3 & -   & 1.02 \\
      dK92 M3 & $M_2$ random from IMF & 1.0 & -   & 0.72 \\
      dK92 M4 & $M_2$ random from IMF & 0.3 & -   & 1.17 \\
      H95 S4  & $\mathrm{d}N\propto\mathrm{d}q$ & 1.0 & 0.0 & 1.66 \\
      H95 S6  & $\mathrm{d}N\propto\mathrm{d}q$ & 1.0 & 1.0 & 2.54 \\
      \hline
   \end{tabular}
   \caption{List of models used in comparison to the observed distribution with their adopted initial mass ratio distributions (IMRD), $\alpha_{CE}$ and $\alpha_{th}$ values. The result of a reduced $\chi^2$ fit to the observed period distribution is also given. Models prefixed with dK92 and H95 are from de Kool (1992) and Han et al. (1995), respectively.
   }
   \label{tab:models}
\end{table}
The models were normalised such that their value equals four at log $P=-0.5$ (similar to DM08) and reduced $\chi^2$ values were subsequently calculated to assist comparison with the observed distribution (Tab. \ref{tab:models}). An alternative normalisation of the models such that the maxima were equal to four was also made, but the $\chi^2$ results were not sufficiently different to change our analysis. With $\chi^2\la1$, dK92 M3 is the best fit ($\sim$80\% probability), whereas dK92 M2 ($\sim$45\%) and dK92 M4 ($\sim$22\%) are also acceptable despite the nonaligned maxima between the observed and model distributions. The disagreement is significant for dK92 M1 and H95 S4 ($\sim$2\%) and highly significant for H95 S6 ($<$0.05\%). 

Realising dK92 M3 as the best fit model strengthens considerably earlier suspicions that prefer a random IMRD based on a much smaller sample than that used here (de Kool 1990; de Kool \& Ritter 2003). A random IMRD is supported by the observed predominance of late-type secondaries (DM08) that is predicted by equivalent models (see Fig. 10c, de Kool \& Ritter 1993; Politano \& Weiler 2006) and the secondary mass distribution of spectroscopic binaries studied by Goldberg et al. (2003).
We refrain from drawing any strong conclusions based on our comparison, but nevertheless it is concerning that \emph{the majority of literature models assume a constant IMRD that predicts substantially more binaries at longer periods}. 
A reliance on these models has created uncertainty regarding whether the large discrepancy for log $P\ga0$ is intrinsically real or caused by selection effects (DM08). DM08 compared the then known period distribution with the H95 S4 and S6 models to suggest a severe inadequacy in all CE population synthesis models. Our larger sample and usage of the best fit dK92 M3 reduces the severe discrepancy to only a \emph{modest} one. The more modest discrepancy is now more easily explained by selection effects rather than a severe flaw in our understanding of CE evolution suggested by DM08. 

At very short periods (log $P\la-1$) the models are in good agreement with observations and any deficiency could be attributed to the sampling frequency of each field (see Sec. \ref{sec:binfrac}). In assessing the contribution of selection effects towards the long period discrepancy (log $P\ga0$), a very useful comparison can be made with the observed period distribution of post-CE white dwarf/main sequence (WDMS) binaries (Rebassa-Mansergas et al. 2008, hereafter RM08). 
In spite of the small samples and differences in selection effects associated with the discovery method of post-CE WDMS (spectroscopic and photometric, Schreiber \& G\"ansicke 2003) and CSPN (photometric) binaries, \emph{the distributions bear a very strong resemblance to each other which strongly suggests they are an acceptable depiction of the true post-CE binary population}. 
RM08 reach the same conclusion based on Monte Carlo simulations that showed they should have been sensitive to detecting longer period binaries in their search. Similarly, DM08 claimed the then known CSPN binaries were also representative of the true population, despite serious misgivings about the contribution of selection effects to the sample discovery. DM08 supported their claim by conducting a theoretical investigation into the expected amplitudes of irradiated binaries for periods $P\ga3$ days and concluded many systems should have been detected by B00 down to amplitudes of $\sim$0.1 mag. 

If the equivalence of the CSPN and post-CE WDMS period distributions is indeed indicative of the true post-CE population, then selection effects alone may not account for the modest discrepancy at longer periods and a revision of the population synthesis models may be required. Alternatively, the most likely cause of the discrepancy is the $I\sim20$ magnitude limit of OGLE-III. At 8 kpc we are likely insensitive to the detection of M-dwarf secondaries at $I\ga21$ (Kirkpatrick \& McCarthy 1994), but irradiation effects may permit detection of some systems. Some secondaries are expected to have spectral types later than M (de Kool \& Ritter 1993; Politano \& Weiler 2006), but these are likely to be rare (Farihi, Becklin \& Zuckerman 2005). A dominant bright primary may also preclude faint secondaries from being detected in some cases. Some more non-detections could be non-irradiated eclipsing binaries, i.e. if M 3-16 were observed at low--moderate inclinations, however these binaries would not necessarily be restricted to log $P\ga0$. A stronger test of the discrepancy could come from RV monitoring surveys that target the longer period regime up to tens of days (M\'endez 1989; De Marco et al. 2004; Sorenson \& Pollacco 2004; Af{\v s}ar \& Bond 2005), but the results from these surveys are largely inconclusive with few, if any, firm periods found as the target sampling seems far from ideal and the typical number of 10--20 observations per object is still quite small. Indeed, the high levels of variability seen (excluding M\'endez 1989) could be attributed to winds and pulsations from the central star alone, rather than being caused by binary companion. Surveys with more frequent sampling may be able to fill the modest discrepancy at longer periods, but we suspect the number of binaries with periods of a few days would remain small based on the post-CE WDMS distribution which includes binaries found using this method. 

The inital investigation into expected variability amplitudes by DM08 raises perhaps the most important selection effect of sensitivity. Around 30\% of OGLE-III binaries exhibit amplitudes $\la0.1$ mag which is markedly different from the known binaries where only Sp 1 has such an amplitude. Although we have considerably improved sensitivity compared to B00 in the low amplitude regime, the fact that we still find a large absence of binaries at longer periods reinforces the representative nature of the current period distribution. Alternatively, some low amplitude binaries may be lost in the larger intrinsic scatter in some fields (Sec. \ref{sec:systematic}). Further investigations following DM08 into the expected low-amplitude population for a variety of irradiated binary configurations and inclinations would be beneficial. 

Constraining $\alpha_{CE}$ is more problematic than constraining the IMRD.
Even though $\alpha_{CE}=1.0$ was used in calculating the best fitting dK92 M3, we caution against adopting $\alpha_{CE}=1.0$ from this work, mainly because different definitions of $\alpha_{CE}$ in the literature complicate comparisons between different models (Livio 1996) and suspicions that $\alpha_{CE}$ may not be constant (Livio 1996; Politano \& Weiler 2007). Furthermore, measuring a post-maximum gradient from the observed period distribution is still hindered by the small sample size even with the new OGLE discoveries.

\subsection{Binary Fraction}
\label{sec:binfrac}
We are now able to derive the first independent estimate of the short period binary CSPN fraction since B00. However, not all OGLE-III fields are sampled uniformly and this can unduly bias any estimate of the binary fraction.  Understanding the capacity of each field to detect short binary CSPN periods ($P\la1$ d) is required before any less sensitive fields can be identified and removed. Usually the Nyquist frequency would be used but this is not necessarily defined for each field that has been sampled intermittently over the course of each Bulge season and many years. We therefore calculated the number of consecutive observations per field, $N(\Delta t_\mathrm{lo})$ and $N(\Delta t_\mathrm{med})$, for separation intervals $\Delta t_\mathrm{lo} < 0.5$ d and $0.5$ d $ < \Delta t_\mathrm{med} < 1.5$ d, respectively. These intervals were chosen because their Nyquist frequencies lied within our desired detection range. Sensitivity to longer periods up to hundreds of days is provided by the long time baseline of the data and because of limited sensitivity to $P\la0.1$ d our search was only conducted for longer periods.

\begin{figure}
   \begin{center}
      \includegraphics[scale=0.35,angle=270]{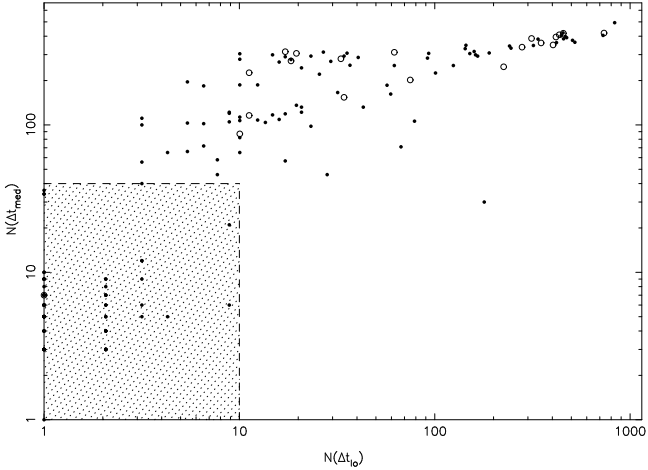}
   \end{center}
   \caption{The relation between the number of consecutive observations $N(\Delta t_\mathrm{lo})$ and $N(\Delta t_\mathrm{med})$ of OGLE-III fields, where $\Delta t_\mathrm{lo} < 0.5$ d and $0.5$ d $<$ $\Delta t_\mathrm{med} < 1.5$ d, respectively. Fields that contain the newly discovered OGLE-III binaries from Fig. \ref{fig:periodic} (open circles) are located within a locus of well-sampled fields (top). The shaded region at lower left contains fields excluded from binary fraction calculations.} 
   \label{fig:ndeltat}
\end{figure}

\begin{figure}
   \centering
      \includegraphics[scale=0.36,angle=270]{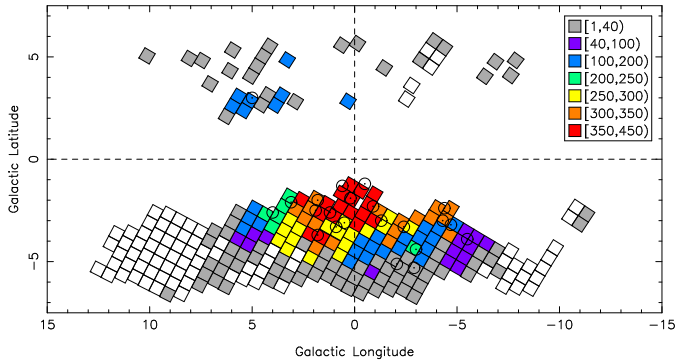}
   \caption{Same as Fig. \ref{fig:survey} but with PNe replaced by binaries from Tab. \ref{tab:periodic} (circled points) and colours indicative of the number of consecutive observations $N(\Delta t_\mathrm{med})$ separated by the interval $0.5$ d $ < \Delta t_\mathrm{med} < 1.5$ d. Less well-sampled fields coloured grey are excluded from binary fraction calculations.}
   \label{fig:sampling}
\end{figure}

Figure \ref{fig:ndeltat} displays the relation between $N(\Delta t_\mathrm{lo})$ and $N(\Delta t_\mathrm{med})$ for all OGLE-III fields. Note the excellent sampling overall and that the new binaries lie only in well-sampled fields along the main locus. The data suggest fields located within the rather conservatively chosen shaded region be excluded from binary fraction calculations (Fig. \ref{fig:ndeltat}). This region contains the field that observed BMP 1800$-$3408 which we consider to be an anomalous detection. Figure \ref{fig:sampling} shows the spatial distribution of $N(\Delta t_\mathrm{med})$ amongst OGLE-III fields with new binaries overlaid. Note the strong concentration towards better sampled southern latitude fields with Pe 1-9 being the only binary detected at positive latitudes.

To estimate the binary fraction we make cuts to Tab. \ref{tab:super} by first removing the PNe belonging to fields largely insensitive to binary CSPN detection. We further remove PNe with remarks `NULL', `NEB' or `P' as we are not able to examine the variability or otherwise of these objects (Sec. \ref{sec:cspnid}). These removed objects contribute towards a `rejected' sample that may include objects we suspect of being affected by systematic effects labelled `S(?)'. This leaves two samples, namely the smaller sample containing only `true' CSPN identifications (T) and a larger sample containing `true' and `likely' CSPN identifications (T$+$L). Some binaries located in fainter PNe fall into the `likely' category in line with other faint PNe and are not included in the T sample. The T$+$L sample constitutes a lower limit to the binary fraction. Note also that we exclude M 1-34 given its quite different nature to the short period binaries. The binary fraction is calculated as the number of binaries (BIN) present in each sample with respect to the total number of objects that were not rejected.

Table \ref{tab:bfraction} contains the calculated binary fractions for both T and T$+$L samples with and without the 10 PNe affected by systematic effects. Our close binary fraction of 12--21\% is in good agreement with 10--15\% found by B00 and 12--33\% determined by Frew \& Parker (2009). However, our sample sizes of $\sim$65 PNe (T) and $\sim$145 (T$+$L) PNe only allow for a meaningful comparison with the B00 sample of $\sim$100 PNe. The early results from Frew \& Parker (2009) are based on a sample of only 33 PNe selected from a larger catalogue of nearby PNe (Frew 2008) which has considerable overlap with the B00 sample. Our estimate also agrees well with model predictions (e.g. $\sim$17\%, Han et al. 1995) and with the fraction of post-CE WDMS binaries (e.g. $\sim$15\%, RM08). Accounting for our insensitivity to M-type secondaries (Sec. \ref{sec:pdist}) could lead to an upward revision of our estimate to $\sim$35\%, but this is left for more sensitive future surveys to confirm. If we include fields removed on the basis of our sampling cut, then this slightly reduces our derived binary fraction to $\sim$10--18\%. The large range in our estimate is caused by the lack of OGLE-III $V$-band data that would have otherwise enabled more `likely' identifications to become `true' via appropriate colour selections. However, we note the lower limit is considered quite robust given that all CSPN candidates for the `likely' sample were examined for variability. 

\begin{table}
   \centering
   \begin{tabular}{lrrrrr}
      \hline\hline
      Sample & S & Passed & Rejected & Binaries & Fraction (\%)\\
      \hline
      T      & N  & 61 & 149 & 13 & 21.3 \\
      T$+$L  & N  & 139 & 149 & 18 & 12.9\\ 
      T      & Y  & 71 & 139 & 13 & 18.3 \\
      T$+$L  & Y  & 149 & 139 & 18 & 12.1\\ 
      \hline
   \end{tabular}
   \caption{Binary fraction calculations for the two samples. The `S' column indicates whether PNe affected by systematic effects are included. See text for further details.} 
   \label{tab:bfraction}
\end{table}
   
\section{Conclusion}
\label{sec:conclusion}
A large catalogue of $\sim$300 Galactic Bulge PNe was constructed after removing 24 new symbiotic stars with the assistance of deep optical spectroscopy. Time series photometry of identified central stars were extracted from the OGLE-III survey and analysed for periodic and aperiodic variability. Nebular contamination and other systematic effects were discussed in the context of often non-stellar PNe in a \emph{stellar-optimised} survey. Images and lightcurves of 21 new periodic close binaries and 1 new likely aperiodic binary were presented, \emph{more than doubling the number of known close binary CSPN}. 

The observed period distribution was compared with predictions made by CE population synthesis models. A best fit was achieved for model 3 from de Kool (1992), which adopted $\alpha_{CE}=1.0$ and a random IMRD. Only a modest discrepancy between the best fit model and the observed distribution is found at longer periods, contrary to most other models that produce a more severe discrepancy. Further comparison with the post-CE WDMS binary period distribution suggests the CSPN and post-CE WDMS binary period distributions are an adequate depiction of the real post-CE population (i.e. binaries with periods longer than about one day are rare). If the discrepancy cannot be resolved by photometric surveys more sensitive to M- and L-type secondaries or RV surveys of large numbers of CSPN conducted with more frequent sampling, then a revision of the CE population synthesis models may be required. 

We have firmly established the close binary fraction to be 10--20\% from our larger and entirely independent sample of binary PNe. This agrees well with previous photometric (10--15\%, B00) and CE population synthesis models (17\%, Han et al. 1995). RV monitoring programs have proved rather unreliable and inconclusive, however our results agree well with an estimate of 15\% by M\'endez (1989). Our estimate does not support the hypothesis that most PNe derive from binaries (Moe \& De Marco 2006), nor the high levels of variability mistaken for binarity in RV monitoring surveys (De Marco et al. 2004; Sorenson \& Pollacco 2004; Af{\v s}ar \& Bond 2005). With the close binary fraction now well determined, more effort is required at the longer period end through detailed study of CSPN showing composite spectra (e.g. Sorenson \& Pollacco 2004), resolved binaries (Ciardullo et al. 1999) and infra-red excesses (e.g. Frew \& Parker 2009). As a shaping mechanism of nebular morphologies close binaries seem to play an important but not major role. Indeed, it has been suggested that the largest role may be fulfilled by low-mass or sub-stellar secondaries (Soker 1997) and the aforementioned studies on longer period binaries will help in this area. This important topic will be further discussed in a following paper with an examination of nebular morphologies of our new sample.

Emphasis should now be placed more on radial velocity monitoring programs and an AAOmega survey of brighter CSPN covered by OGLE-III would prove particularly useful in better understanding the selection effects of our survey. However, there are still advancements to be made via photometric campaigns. A bias against the brightest PNe in OGLE-III requires dedictated independent followup perhaps using narrow-band filters to remove nebular contamination. Planned surveys such as Vista Variables in the Via Lactea (VVV, see Arnaboldi et al. 2007) are likely to find more binaries outside OGLE-III fields \emph{provided} field sampling is not too heavily biased towards higher stellar density fields already covered by OGLE-III. Since VVV is conducted in the $K_s$-band the survey may be \emph{more} sensitive to binaries with later-type secondaries than OGLE-III and would therefore provide greater understanding of the selection effects of our survey. Dedicated photometric surveys with telescopes such as SkyMapper (Keller et al. 2007) are required to survey regions outside microlensing surveys or where the sampling is not always ideal and such surveys would prove very fruitful indeed in monitoring the now very large population of Galactic Bulge and Southern Galactic Plane PNe (Parker et al. 2006; Miszalski et al. 2008a).

\begin{acknowledgements}
    BM acknowledges the support of an Australian Postgraduate Award and further support from Macquarie University, Strasbourg Observatory, PICS and European Southern Observatory. AFJM is grateful for financial assistance to NSERC (Canada) and FQRNT (Qu\'ebec). 
    Discussions with Simon O'Toole, David Frew and Paul Dobbie are greatly appreciated.
    The bulk of the AAOmega observations were deftly conducted by Paul Dobbie and special thanks go to Stephen Fine and Scott Croom for additional AAOmega observations. Rob Sharp is acknowledged for his expertise and assistance with 2dfdr that was used to reduce the AAOmega data. Thanks to Anna Kovacevic for inspection of some [OIII] images of PNe in our sample and for taking the H$\alpha$ image of K 6-34. 
    AA and BM thank Claudio Melo for his excellent support and company during our stay at Paranal. We thank the Australian National University and European Southern Observatory telescope time allocation committees. We acknowledge helpful suggestions from the referee, Orsola De Marco, that improved this paper.
         The OGLE project is partially supported by the Polish MNiSW grant N20303032/4275.
    This research has made use of SAOImage \textsc{ds9}, developed by Smithsonian Astrophysical Observatory, the SIMBAD data base operated at CDS, Strasbourg, France, and makes use of data products from the 2MASS survey, which is a joint project of the University of Massachusetts and the Infrared Processing and Analysis Centre/California Institute of Technology, funded by the National Aeronautical and Space Administration and the National Science Foundation.
    Images were presented from observations obtained at the Gemini Observatory, which is operated by the
    Association of Universities for Research in Astronomy, Inc., under a cooperative agreement
    with the NSF on behalf of the Gemini partnership: the National Science Foundation (United
    States), the Science and Technology Facilities Council (United Kingdom), the
    National Research Council (Canada), CONICYT (Chile), the Australian Research Council
    (Australia), Minisrio da Ciência e Tecnologia (Brazil) and SECYT (Argentina).
\end{acknowledgements}

\clearpage \onecolumn
\begin{appendix}
   \section{List of Planetary Nebulae}
\begin{center}
   \tablehead{
   \hline
   \hline
   Field & PN G & Name & RA (J2000) & DEC (J2000) & Remarks            \\
   \hline}
   \tabletail{
   \hline
   }
   \topcaption{Planetary nebulae with OGLE-III coverage. Remarks are explained in the text.}
\begin{supertabular}{llllll}
   \label{tab:super}
BLG366          & 352.1$+$05.1 & M 2-8 & 17 05 30.7 & $-$32 32 08 & T                                \\  
BLG363          & 355.6$+$05.1 & MPA 1714$-$2946 & 17 14 48.2 & $-$29 46 47 & T,B                     \\  
BLG358          & 356.8$+$05.1 & PHR 1717$-$2845 & 17 17 48.7 & $-$28 45 57 & T,B                    \\  
BLG358          & 356.4$+$04.8 & PHR 1718$-$2914 & 17 18 12.2 & $-$29 14 58 & P                      \\  
BLG357          & 359.8$+$05.6 & M 2-12 & 17 24 01.4 & $-$25 59 23         & T                       \\  
BLG357          & 000.1$+$05.7 & PHR 1724$-$2543 & 17 24 04.3 & $-$25 43 14 & P                       \\  
BLG355          & 000.3$+$05.6 & PHR 1725$-$2534 & 17 25 13.2 & $-$25 34 16 & P                       \\  
BLG200          & 348.7$-$02.5 & MPA 1726$-$3950 & 17 26 34.3 & $-$39 50 20 & L                     \\  
BLG333          & 000.1$+$02.6 & Al 2-J & 17 35 35.5 & $-$27 24 05         & L                        \\  
BLG336          & 004.4$+$05.3 & K 6-27 & 17 35 53.8 & $-$22 20 02         & P           \\  
BLG333          & 000.5$+$02.8 & PHR 1735$-$2659 & 17 35 55.7 & $-$26 59 17 & L            \\  
BLG352          & 006.0$+$05.6 & PHR 1738$-$2052 & 17 38 25.7 & $-$20 52 18 & T,B            \\  
BLG352          & 005.8$+$05.1 & H 2-16 & 17 39 55.4 & $-$21 14 11         & L            \\  
BLG337          & 005.0$+$04.4 & H 1-27 & 17 40 18.0 & $-$22 19 18         & NULL           \\
BLG338          & 005.2$+$04.2 & M 3-13 & 17 41 36.7 & $-$22 13 03         & T,S?           \\  
BLG334          & 003.1$+$02.9 & Hb 4 & 17 41 52.8 & $-$24 42 08           & T,S           \\  
BLG346          & 003.6$+$03.1 & M 2-14 & 17 41 57.4 & $-$24 11 16         & NULL           \\
BLG338          & 005.4$+$04.0 & PHR 1742$-$2214 & 17 42 54.7 & $-$22 14 16 & L            \\  
BLG348          & 007.1$+$04.9 & PHR 1743$-$2013 & 17 43 34.1 & $-$20 13 56 & NULL           \\
BLG347          & 003.5$+$02.6 & PHR 1743$-$2431 & 17 43 39.4 & $-$24 31 53 & T,B            \\  
BLG347          & 003.6$+$02.7 & PHR 1743$-$2424 & 17 43 49.7 & $-$24 24 06 & P            \\  
BLG129          & 355.2$-$02.5 & H 1-29 & 17 44 13.9 & $-$34 17 34         & NULL           \\
BLG129          & 355.4$-$02.4 & M 3-14 & 17 44 20.6 & $-$34 06 40         & NULL           \\
BLG129          & 355.6$-$02.3 & PHR 1744$-$3355 & 17 44 27.8 & $-$33 55 20 & T,B,BIN           \\  
BLG347          & 003.9$+$02.6 & K 5-14 & 17 44 32.6 & $-$24 13 27         & T,S           \\  
BLG347          & 004.0$+$02.6 & PHR 1744$-$2406 & 17 44 46.3 & $-$24 06 59 & L \\  
BLG129          & 355.4$-$02.6 & PHR 1745$-$3413 & 17 45 03.6 & $-$34 13 26 & L            \\  
BLG342          & 005.1$+$03.2 & PHR 1745$-$2254 & 17 45 17.0 & $-$22 54 58 & T,B            \\  
BLG112          & 353.8$-$03.7 & PHR 1745$-$3609 & 17 45 32.2 & $-$36 09 57 & T,B            \\  
BLG121          & 355.1$-$02.9 & H 1-31 & 17 45 32.4 & $-$34 33 56         & T,S            \\  
BLG342          & 005.0$+$03.0 & Pe 1-9 & 17 45 36.7 & $-$23 02 26         & T,B,BIN\\
BLG130          & 355.6$-$02.7 & H 1-32 & 17 46 06.2 & $-$34 03 46         & NULL           \\
BLG106          & 353.0$-$04.4 & PHR 1746$-$3713 & 17 46 24.5 & $-$37 13 09 & P            \\  
BLG106          & 353.0$-$04.4a & MPA 1746$-$3712 & 17 46 25.3 & $-$37 12 48& T,B            \\  
BLG130          & 355.4$-$03.1 & PPA 1746$-$3428 & 17 46 51.4 & $-$34 28 01 & L             \\  
BLG106          & 352.8$-$04.6 & PPA 1746$-$3725 & 17 46 59.8 & $-$37 25 36 & L            \\  
BLG139          & 356.1$-$02.7 & PPA 1747$-$3341 & 17 47 04.8 & $-$33 41 03 & NEB            \\  
BLG121          & 355.3$-$03.2 & PPA 1747$-$3435 & 17 47 08.4 & $-$34 35 43 & T,B,BIN            \\  
BLG139          & 356.3$-$02.6 & MPA 1747$-$3326 & 17 47 27.6 & $-$33 26 38 & NONE            \\  
BLG342          & 005.1$+$02.6 & PHR 1747$-$2311 & 17 47 30.7 & $-$23 11 49 & T,B            \\  
BLG130          & 355.7$-$03.0 & H 1-33 & 17 47 49.4 & $-$34 08 05         & T,BIN            \\  
BLG109          & 353.4$-$04.5 & K 6-13 & 17 48 01.0 & $-$36 50 08         & L            \\  
BLG339          & 005.5$+$02.7 & H 1-34 & 17 48 07.7 & $-$22 46 47         & NULL           \\
BLG343          & 005.3$+$02.5 & - & 17 48 12.7 & $-$22 59 39              & T            \\  
BLG117          & 354.5$-$03.9 & Sab 41         & 17 48 16.3 & $-$35 38 31 & T,B,BIN            \\  
BLG340          & 006.0$+$02.9 & - & 17 48 25.0 & $-$22 11 52              & T            \\  
BLG117,122 & 355.0$-$03.7 & K 5-18 & 17 48 29.5 & $-$35 05 29         & P            \\  
BLG340          & 006.0$+$02.8 & Th 4-3 & 17 48 37.4 & $-$22 16 49         & T            \\  
BLG117          & 354.7$-$03.9 & MPA 1748$-$3530 & 17 48 48.6 & $-$35 30 30 & L            \\  
BLG131          & 355.7$-$03.4 & H 2-23 & 17 48 58.1 & $-$34 21 54         & NULL           \\
BLG131          & 355.7$-$03.5 & H 1-35 & 17 49 13.9 & $-$34 22 53         & NULL           \\
BLG179          & 358.9$-$01.5 & JaSt 65 & 17 49 19.9 & $-$30 36 06        & T,S           \\  
BLG139,140 & 356.2$-$03.2 & PHR 1749$-$3347 & 17 49 23.8 & $-$33 47 41 & L            \\  
BLG131          & 356.1$-$03.3 & H 2-26 & 17 49 50.9 & $-$34 00 31         & L            \\  
BLG179          & 359.0$-$01.6 & GLMP 647 & 17 49 52.6 & $-$30 33 02       & T            \\  
BLG194          & 000.1$-$01.0 & JaSt 69 & 17 50 10.1 & $-$29 19 05        & P            \\  
BLG147          & 356.8$-$03.0 & K 5-20 & 17 50 10.8 & $-$33 14 18         & T            \\  
BLG179          & 359.1$-$01.7 & M 1-29 & 17 50 18.0 & $-$30 34 54         & NEB            \\  
BLG194          & 000.1$-$01.1 & M 3-43 & 17 50 24.2 & $-$29 25 19         & T            \\  
BLG107          & 353.2$-$05.2 & H 1-38 & 17 50 45.4 & $-$37 23 53         & L            \\  
BLG100          & 359.7$-$01.4 & JaSt 73 & 17 50 47.8 & $-$29 53 14        & T            \\  
BLG194          & 000.1$-$01.2 & JaSt 75 & 17 50 48.0 & $-$29 24 43        & NULL           \\
BLG114          & 354.6$-$04.5 & PPA 1750$-$3548 & 17 50 56.2 & $-$35 48 49 & L            \\  
BLG123          & 355.3$-$04.1 & PHR 1750$-$3500 & 17 50 56.9 & $-$35 00 46 & NULL          \\
BLG194          & 000.0$-$01.3 & PPA 1751$-$2933 & 17 51 00.5 & $-$29 33 51 & L             \\  
BLG132          & 356.2$-$03.6 & PPA 1751$-$3401 & 17 51 07.0 & $-$34 01 41 & L            \\  
BLG123          & 355.4$-$04.0 & Pe 1-10 & 17 51 12.2 & $-$34 55 22        & T            \\  
BLG179,180 & 359.3$-$01.8 & M 3-44 & 17 51 19.0 & $-$30 23 53         & T            \\  
BLG140          & 356.5$-$03.4 & MPA 1751$-$3339 & 17 51 20.6 & $-$33 39 13 & P            \\  
BLG118          & 354.9$-$04.4 & PHR 1751$-$3531 & 17 51 23.0 & $-$35 31 18 & L            \\  
BLG171          & 358.9$-$02.1 & PHR 1751$-$3059 & 17 51 38.9 & $-$30 59 54 & L            \\  
BLG341          & 006.3$+$02.2 & MPA 1751$-$2223 & 17 51 40.0 & $-$22 23 18 & L            \\  
BLG100          & 359.5$-$01.8 & PHR 1751$-$3012 & 17 51 44.2 & $-$30 12 47 & P            \\  
BLG163          & 358.3$-$02.5 & M 4-7 & 17 51 44.6 & $-$31 36 01          & T            \\  
BLG100          & 359.7$-$01.7 & K 6-15 & 17 51 48.7 & $-$30 02 34         & T            \\  
BLG140          & 356.5$-$03.6 & H 2-27 & 17 51 50.6 & $-$33 47 36         & L            \\  
BLG171          & 358.6$-$02.4 & K 6-16 & 17 52 00.2 & $-$31 17 50         & T            \\  
BLG148          & 357.4$-$03.1 & PHR 1752$-$3244 & 17 52 00.7 & $-$32 44 08 & L            \\  
BLG194          & 000.4$-$01.3 & JaSt 2-8 & 17 52 03.8 & $-$29 16 42       & L,B           \\  
BLG100          & 359.7$-$01.8 & M 3-45 & 17 52 06.0 & $-$30 05 14         & T            \\  
BLG155          & 357.6$-$03.0 & PHR 1752$-$3233 & 17 52 11.8 & $-$32 33 08 & T            \\  
BLG155          & 357.6$-$03.0a & PHR 1752$-$3230 & 17 52 16.1 & $-$32 30 07& T,B            \\  
BLG123          & 355.4$-$04.3 & K 5-34 & 17 52 23.0 & $-$35 04 18         & L             \\  
BLG141          & 356.8$-$03.6 & PHR 1752$-$3330 & 17 52 29.2 & $-$33 30 04 & T          \\  
BLG148          & 357.4$-$03.2 & M 2-16 & 17 52 34.3 & $-$32 45 51         & NULL           \\
BLG195          & 000.6$-$01.3 & Bl 3-15 & 17 52 36.0 & $-$29 06 39        & T,BIN            \\  
BLG171          & 358.7$-$02.5 & PHR 1752$-$3116 & 17 52 36.5 & $-$31 16 27 & NONE            \\  
BLG341          & 006.4$+$02.0 & M 1-31 & 17 52 41.5 & $-$22 21 57         & NULL           \\
BLG101          & 000.0$-$01.8 & JaSt 83 & 17 52 45.1 & $-$29 50 59        & L            \\  
BLG180          & 359.1$-$02.3 & M 3-16 & 17 52 46.1 & $-$30 49 35         & T,BIN            \\  
BLG195          & 000.5$-$01.5 & JaSt 2-9 & 17 52 47.8 & $-$29 17 23       & L             \\  
BLG101          & 000.1$-$01.7 & PHR 1752$-$2941 & 17 52 49.0 & $-$29 41 59 & P            \\  
BLG101          & 359.9$-$01.8 & MPA 1752$-$2953 & 17 52 49.2 & $-$29 53 01 & P            \\  
BLG195          & 000.3$-$01.6 & PHR 1752$-$2930 & 17 52 52.1 & $-$29 30 01 & P            \\  
BLG101          & 000.0$-$01.8a & PHR 1752$-$2953 & 17 52 58.3 & $-$29 53 23& P           \\  
BLG123          & 355.9$-$04.2 & M 1-30 & 17 52 59.0 & $-$34 38 23         & NULL           \\
BLG195          & 000.6$-$01.4 & PHR 1753$-$2905 & 17 53 00.7 & $-$29 05 53 & L            \\  
BLG132          & 356.0$-$04.2 & PHR 1753$-$3428 & 17 53 04.8 & $-$34 28 39 & L            \\  
BLG180          & 359.3$-$02.3 & PHR 1753$-$3038 & 17 53 16.3 & $-$30 38 40 & L            \\  
BLG155          & 357.6$-$03.3 & H 2-29 & 17 53 16.8 & $-$32 40 38         & T,B,BIN            \\  
BLG132          & 356.5$-$03.9 & H 1-39 & 17 53 21.1 & $-$33 55 58         & NULL           \\
BLG195          & 000.5$-$01.6 & Al 2-Q & 17 53 25.2 & $-$29 17 08         & L            \\  
BLG195          & 000.7$-$01.5 & JaSt 2-11 & 17 53 26.9 & $-$29 08 16      & L            \\  
BLG171          & 358.7$-$02.7 & Al 2-R & 17 53 36.2 & $-$31 25 25         & NEB            \\  
BLG148          & 357.4$-$03.5 & M 2-18 & 17 53 37.9 & $-$32 58 48         & NULL           \\
BLG180          & 359.2$-$02.4 & PHR 1753$-$3051 & 17 53 39.8 & $-$30 51 25 & L             \\  
BLG124          & 355.9$-$04.4 & PHR 1753$-$3443 & 17 53 40.3 & $-$34 43 41 & T,B,S?           \\  
BLG101          & 000.2$-$01.9 & M 2-19 & 17 53 45.6 & $-$29 43 46         & T,BIN            \\  
BLG195          & 000.8$-$01.5 & Bl O & 17 53 49.7 & $-$28 59 11           & T,S            \\  
BLG156          & 357.8$-$03.3 & PHR 1753$-$3228 & 17 53 55.9 & $-$32 28 50 & P           \\  
BLG195          & 000.5$-$01.7 & JaSt 96 & 17 53 57.6 & $-$29 20 14        & P            \\  
BLG180,181 & 359.7$-$02.2 & PPA 1753$-$3021 & 17 53 59.3 & $-$30 21 49 & L            \\  
BLG133          & 356.5$-$04.1 & PPA 1754$-$3358 & 17 54 03.1 & $-$33 58 51 & NULL           \\
BLG101          & 000.0$-$02.1 & MPA 1754$-$2957 & 17 54 04.3 & $-$29 57 27 & L            \\  
BLG119          & 355.5$-$04.8 & PHR 1754$-$3515 & 17 54 17.5 & $-$35 15 39 & L            \\  
BLG101          & 000.4$-$01.9 & M 2-20 & 17 54 25.4 & $-$29 36 09         & T,S?           \\  
BLG119          & 355.2$-$05.0 & PHR 1754$-$3533 & 17 54 30.7 & $-$35 33 08 & L            \\  
BLG133          & 356.2$-$04.4 & Cn 2-1 & 17 54 32.9 & $-$34 22 22         & NULL           \\
BLG172          & 358.7$-$03.0 & K 6-34 & 17 54 41.3 & $-$31 31 43         & T,B,BIN            \\  
BLG172          & 359.1$-$02.9 & M 3$-$46 & 17 55 05.5 & $-$31 12 17       & NEB            \\  
BLG172          & 359.3$-$02.8 & MPA 1755$-$3058 & 17 55 13.1 & $-$30 58 13 & L            \\  
BLG103          & 000.2$-$02.3 & Bl 3-10 & 17 55 20.6 & $-$29 57 36        & NEB            \\  
BLG181          & 359.7$-$02.6 & H 1-40 & 17 55 36.0 & $-$30 33 33         & T            \\  
BLG103          & 000.0$-$02.5 & K 6-36 & 17 55 52.8 & $-$30 15 41         & T            \\  
BLG164          & 358.8$-$03.3 & PK 358$-$033 & 17 56 02.4 & $-$31 38 23   & T            \\  
BLG205          & 000.9$-$02.0 & Bl 3-13 & 17 56 02.6 & $-$29 11 16        & NULL           \\
BLG103          & 359.9$-$02.6 & PHR 1756$-$3019 & 17 56 06.5 & $-$30 19 36 & P            \\  
BLG156          & 357.9$-$03.8 & H 2-30 & 17 56 13.9 & $-$32 37 22         & T            \\  
BLG172          & 359.3$-$03.1 & M 3-17 & 17 56 25.7 & $-$31 04 17         & T,S?           \\  
BLG172          & 359.2$-$03.1 & PHR 1756$-$3112 & 17 56 28.8 & $-$31 12 35 & L             \\  
BLG214          & 001.5$-$01.8 & JaSt 2-19 & 17 56 33.8 & $-$28 30 30      & L            \\  
BLG205          & 001.2$-$02.0 & PHR 1756$-$2857 & 17 56 36.5 & $-$28 57 18 & P            \\  
BLG142          & 357.0$-$04.4 & PHR 1756$-$3342 & 17 56 39.6 & $-$33 42 31 & T,B,BIN            \\  
BLG164          & 358.6$-$03.6 & PHR 1756$-$3157 & 17 56 43.7 & $-$31 57 33 & P            \\  
BLG134          & 356.6$-$04.7 & PHR 1756$-$3414 & 17 56 48.2 & $-$34 14 32 & L            \\  
BLG134          & 356.7$-$04.7 & MPA 1757$-$3410 & 17 57 00.7 & $-$34 10 41 & T,B            \\  
BLG103,182 & 000.0$-$02.9 & MPA 1757$-$3021 & 17 57 14.2 & $-$30 21 53 &L          \\
BLG134          & 356.7$-$04.8 & H 1-41 & 17 57 19.0 & $-$34 09 49         & T            \\  
BLG142          & 357.2$-$04.5 & H 1-42 & 17 57 25.2 & $-$33 35 43         & T            \\  
BLG214          & 001.8$-$02.0 & PHR 1757$-$2824 & 17 57 42.2 & $-$28 24 07 & T,B,BIN            \\  
BLG188          & 000.3$-$02.8 & M 3-47 & 17 57 43.4 & $-$30 02 30         & T            \\  
BLG165          & 358.9$-$03.6 & PPA 1757$-$3144 & 17 57 48.7 & $-$31 44 05 & P            \\  
BLG173          & 359.4$-$03.3 & PHR 1757$-$3106 & 17 57 51.6 & $-$31 06 11 & L            \\  
BLG104          & 000.7$-$02.7 & M 2-21 & 17 58 09.6 & $-$29 44 20         & NULL           \\
BLG165          & 358.9$-$03.7 & H 1-44 & 17 58 10.6 & $-$31 42 56         & T            \\  
BLG173          & 359.4$-$03.4 & H 2-33 & 17 58 12.5 & $-$31 08 06         & T            \\  
BLG143          & 357.1$-$04.7 & H 1-43 & 17 58 14.4 & $-$33 47 37         & T            \\  
BLG205          & 001.6$-$02.2 & MPA 1758$-$2841 & 17 58 14.8 & $-$28 41 30 & P          \\
BLG188          & 000.4$-$02.9 & M 3-19 & 17 58 19.4 & $-$30 00 40         & T            \\  
BLG150          & 357.8$-$04.4 & PHR 1758$-$3304 & 17 58 25.9 & $-$33 04 59 & L            \\  
BLG223          & 002.6$-$01.7 & PHR 1758$-$2729 & 17 58 28.9 & $-$27 29 40 & P           \\ 
BLG223          & 002.5$-$01.7 & Pe 2-11 & 17 58 31.2 & $-$27 37 06        & T            \\  
BLG143          & 357.4$-$04.6 & M 2-22 & 17 58 32.6 & $-$33 28 37         & T            \\  
BLG223          & 002.3$-$01.9 & PHR 1758$-$2756 & 17 58 35.0 & $-$27 56 56 & P            \\  
BLG157,158 & 358.5$-$04.2 & H 1-46 & 17 59 02.4 & $-$32 21 44         & NULL           \\
BLG104          & 001.2$-$02.6 & PHR 1759$-$2915 & 17 59 02.9 & $-$29 15 01 & L,BIN            \\  
BLG188          & 000.5$-$03.1 & KFL 1 & 17 59 15.6 & $-$30 02 48          & L             \\  
BLG215          & 002.1$-$02.2 & M 3-20 & 17 59 19.4 & $-$28 13 48         & T            \\  
BLG173,174 & 359.5$-$03.7 & MPA 1759$-$3116 & 17 59 25.5 & $-$31 16 54 & P            \\  
BLG188          & 000.5$-$03.1a & MPA 1759$-$3007 & 17 59 25.6 & $-$30 07 15& L,BIN            \\  
BLG232          & 003.1$-$01.6 & PHR 1759$-$2706 & 17 59 26.2 & $-$27 06 34 & NULL           \\
BLG232          & 003.0$-$01.7 & PHR 1759$-$2712 & 17 59 33.1 & $-$27 12 50 & T,B            \\  
BLG143          & 357.7$-$04.8 & BMP 1759$-$3321 & 17 59 45.2 & $-$33 21 13 & T,B            \\  
BLG183          & 000.2$-$03.4 & PHR 1759$-$3030 & 17 59 47.8 & $-$30 30 35 & P            \\  
BLG215          & 001.9$-$02.5 & PPA 1759$-$2834 & 17 59 52.6 & $-$28 34 47 & L,BIN            \\  
BLG151          & 358.0$-$04.6 & Sa 3-107 & 17 59 55.0 & $-$32 59 12       & T            \\  
BLG232          & 003.3$-$01.6 & PHR 1759$-$2651 & 17 59 55.2 & $-$26 51 49 & L \\
BLG232          & 002.9$-$01.8 & MPA 1759$-$2719 & 17 59 55.5 & $-$27 19 17 &P          \\
BLG166          & 359.0$-$04.1 & M 3-48 & 17 59 56.6 & $-$31 54 27         & NEB            \\  
BLG188          & 000.6$-$03.2 & MPA 1759$-$3004 & 17 59 56.7 & $-$30 04 28 & L            \\  
BLG206          & 001.6$-$02.6 & PHR 1759$-$2853 & 17 59 58.8 & $-$28 53 57 & P            \\  
BLG158          & 358.4$-$04.5 & K 6-38 & 17 59 59.0 & $-$32 35 58         & T            \\  
BLG183          & 000.2$-$03.4a& MPA 1800$-$3026& 18 00 00.0 & $-$30 26 08  & P          \\ 
BLG206          & 001.7$-$02.6 & PPA 1800$-$2846 & 18 00 00.7 & $-$28 46 27 & NULL            \\  
BLG183          & 000.3$-$03.4 & MPA 1800$-$3023 & 18 00 11.1 & $-$30 23 49 & T,B            \\  
BLG135          & 356.8$-$05.4 & H 2-35 & 18 00 18.2 & $-$34 27 40         & L             \\  
BLG215          & 002.1$-$02.4 & PPA 1800$-$2818 & 18 00 18.7 & $-$28 18 35 & L            \\  
BLG215          & 002.0$-$02.5 & PPA 1800$-$2826 & 18 00 18.7 & $-$28 26 08 & L            \\  
BLG206          & 001.5$-$02.8 & PPA 1800$-$2904 & 18 00 22.3 & $-$29 04 39 & T            \\  
BLG135          & 357.1$-$05.3 & BMP 1800$-$3408 & 18 00 26.1 & $-$34 08 03 & T,B,BIN           \\  
BLG206          & 001.8$-$02.7 & PHR 1800$-$2842 & 18 00 36.0 & $-$28 42 51 & P            \\  
BLG105          & 001.2$-$03.0 & H 1-47 & 18 00 37.7 & $-$29 21 51         & NULL           \\
BLG232          & 003.4$-$01.8 & PHR 1800$-$2653 & 18 00 42.2 & $-$26 53 37 & L            \\  
BLG144          & 357.5$-$05.1 & PPA 1800$-$3341 & 18 00 43.2 & $-$33 41 59 & T            \\  
BLG215          & 002.2$-$02.5 & KFL 2 & 18 00 59.8 & $-$28 16 19          & L            \\  
BLG224          & 002.8$-$02.2 & Pe 2-12 & 18 01 10.1 & $-$27 38 26        & NEB            \\  
BLG189          & 000.9$-$03.3 & PHR 1801$-$2947 & 18 01 13.4 & $-$29 47 00 & L,BIN             \\  
BLG151          & 357.9$-$05.1 & M 1-34 & 18 01 22.1 & $-$33 17 43         & L,BIN?           \\  
BLG233          & 003.1$-$02.1 & PHR 1801$-$2718 & 18 01 24.5 & $-$27 18 09 & L,BIN            \\  
BLG224          & 002.7$-$02.4 & PPA 1801$-$2746 & 18 01 32.4 & $-$27 46 07 & L            \\  
BLG216          & 002.1$-$02.8 & PHR 1801$-$2831 & 18 01 32.4 & $-$28 31 45 & L            \\  
BLG216          & 002.2$-$02.7 & M 2-23 & 18 01 42.7 & $-$28 25 44         & T,S            \\  
BLG151          & 358.0$-$05.1 & Pe 1-11 & 18 01 42.7 & $-$33 15 25        & NULL \\  
BLG216          & 002.4$-$02.6 & PHR 1801$-$2809 & 18 01 45.4 & $-$28 09 37 & P            \\  
BLG224          & 002.6$-$02.5 & MPA 1801$-$2755 & 18 01 50.1 & $-$27 55 26 & P          \\ 
BLG216          & 002.5$-$02.6 & MPA 1802$-$2803 & 18 02 04.5 & $-$28 03 39 & L          \\
BLG189          & 000.7$-$03.7 & M 3-22 & 18 02 19.2 & $-$30 14 25         & T            \\  
BLG144          & 357.5$-$05.5 & PPA 1802$-$3350 & 18 02 23.8 & $-$33 50 48 & L            \\  
BLG105          & 001.4$-$03.4 & ShWi 1 & 18 02 26.4 & $-$29 25 05         & L             \\  
BLG175          & 359.7$-$04.4a & PPA 1802$-$3124 & 18 02 30.3 & $-$31 24 17& P            \\  
BLG159          & 359.0$-$04.8 & M 2-25 & 18 02 46.6 & $-$32 09 29         & P            \\  
BLG207          & 002.0$-$03.1 & PHR 1802$-$2847 & 18 02 48.5 & $-$28 47 41 & L            \\  
BLG207          & 001.8$-$03.2 & MPA 1802$-$2900 & 18 02 50.2 & $-$29 00 20 & L            \\  
BLG224          & 003.0$-$02.6 & KFL 4 & 18 02 51.4 & $-$27 41 01          & T            \\  
BLG216          & 002.6$-$02.8a & MPA 1802$-$2807 & 18 02 53.6 & $-$28 07 56& L          \\
BLG216          & 002.6$-$02.8 & PHR 1803$-$2804 & 18 03 08.4 & $-$28 04 15 & P            \\  
BLG233          & 003.5$-$02.3 & PHR 1803$-$2702 & 18 03 09.8 & $-$27 02 46 & L           \\  
BLG207          & 002.0$-$03.2 & PHR 1803$-$2848 & 18 03 09.8 & $-$28 48 51 & L             \\  
BLG233          & 003.6$-$02.3 & M 2-26 & 18 03 12.0 & $-$26 58 30         & L \\
BLG159          & 359.0$-$04.9 & PHR 1803$-$3218 & 18 03 16.8 & $-$32 18 26 & L            \\  
BLG184          & 000.3$-$04.2 & MPA 1803$-$3043 & 18 03 21.7 & $-$30 43 36 & P            \\  
BLG216          & 002.4$-$03.1 & PPA 1803$-$2826 & 18 03 24.7 & $-$28 26 25 & NULL           \\
BLG184          & 000.3$-$04.2a & MPA 1803$-$3046 & 18 03 25.8 & $-$30 46 25& P            \\  
BLG159          & 358.7$-$05.1 & SB 53 & 18 03 28.6 & $-$32 37 25          & L            \\  
BLG190          & 001.2$-$03.8 & PHR 1803$-$2947 & 18 03 31.2 & $-$29 47 11 & L              \\  
BLG207          & 002.0$-$03.4 & PPA 1803$-$2855 & 18 03 37.4 & $-$28 55 41 & L            \\  
BLG175          & 359.9$-$04.5 & M 2-27 & 18 03 52.6 & $-$31 17 47         & T,S            \\  
BLG159          & 358.7$-$05.2 & H 1-50 & 18 03 53.5 & $-$32 41 42         & T,S?           \\  
BLG234          & 003.8$-$02.4 & PHR 1804$-$2653 & 18 04 02.6 & $-$26 53 27 & L           \\  
BLG136          & 357.1$-$06.1 & M 3-50 & 18 04 05.3 & $-$34 28 38         & NEB            \\  
BLG216,217 & 002.4$-$03.2 & Sa 3-115 & 18 04 05.5 & $-$28 27 51       & L            \\  
BLG167          & 359.6$-$04.8 & H 2-36 & 18 04 07.7 & $-$31 39 10         & L            \\  
BLG196          & 001.6$-$03.7 & MPA 1804$-$2926 & 18 04 10.9 & $-$29 26 33 & P          \\
BLG217          & 002.6$-$03.1 & PHR 1804$-$2816 & 18 04 13.2 & $-$28 16 12 & P             \\  
BLG152          & 358.4$-$05.5 & PHR 1804$-$3306 & 18 04 14.9 & $-$33 06 30 & T,B            \\  
BLG225          & 002.9$-$03.0 & PHR 1804$-$2757 & 18 04 17.3 & $-$27 57 12 & L             \\  
BLG196          & 001.8$-$03.7 & PHR 1804$-$2913 & 18 04 28.5 & $-$29 13 57   & T,BIN            \\  
BLG208          & 002.3$-$03.4 & H 2-37 & 18 04 28.8 & $-$28 37 38         & T \\
BLG127          & 356.7$-$06.4 & H 1-51 & 18 04 29.3 & $-$34 58 00         & L            \\  
BLG175          & 000.2$-$04.6 & Sa 3-117 & 18 04 44.2 & $-$31 02 49       & P            \\  
BLG190          & 000.9$-$04.2 & PHR 1804$-$3016 & 18 04 48.0 & $-$30 16 49 & L            \\  
BLG217          & 002.4$-$03.4 & PHR 1804$-$2833 & 18 04 55.4 & $-$28 33 08 & P,B            \\  
BLG153          & 358.6$-$05.5 & M 3-51 & 18 04 56.2 & $-$32 54 01         & L            \\  
BLG241          & 004.0$-$02.5 & PHR 1804$-$2642 & 18 04 58.8 & $-$26 42 35 & P            \\  
BLG241          & 004.0$-$02.6 & PHR 1804$-$2645 & 18 04 59.5 & $-$26 45 17 & T,BIN            \\  
BLG176          & 000.3$-$04.6 & M 2-28 & 18 05 02.6 & $-$30 58 17         & NEB           \\  
BLG234          & 003.6$-$02.8 & MPA 1805$-$2712 & 18 05 03.2 & $-$27 12 33 & P            \\  
BLG185          & 000.6$-$04.5 & PM 1-206 & 18 05 12.5 & $-$30 42 13       & T,B            \\  
BLG234          & 003.5$-$02.9 & MPA 1805$-$2721 & 18 05 13.1 & $-$27 21 08 & NULL           \\  
BLG241          & 004.2$-$02.5 & PHR 1805$-$2631 & 18 05 20.2 & $-$26 31 45 & L            \\  
BLG217          & 002.6$-$03.4 & M 1-37 & 18 05 25.7 & $-$28 22 04         & T            \\  
BLG234          & 004.0$-$02.7 & PPA 1805$-$2649 & 18 05 26.4 & $-$26 49 03 & P           \\  
BLG168          & 359.7-05.0 & PHR 1805-3140 & 18 05 30.2 & $-$31 40 16     & L             \\  
BLG217          & 002.9$-$03.3 & PHR 1805$-$2804 & 18 05 44.4 & $-$28 04 46 & L            \\  
BLG208          & 002.1$-$03.8 & MPA 1805$-$2902 & 18 05 55.4 & $-$29 02 46 & P          \\
BLG241          & 004.3$-$02.6 & H 1-53 & 18 05 57.4 & $-$26 29 42         & NULL           \\
BLG234          & 003.6$-$03.0 & PHR 1805$-$2723 & 18 05 57.8 & $-$27 23 09 & L            \\  
BLG208          & 002.4$-$03.7 & M 1-38 & 18 06 05.8 & $-$28 40 30         & T            \\  
BLG128          & 357.3$-$06.5 & SB 50 & 18 06 08.2 & $-$34 33 30          & L            \\  
BLG226          & 003.3$-$03.3 & PHR 1806$-$2747 & 18 06 28.1 & $-$27 47 16 & L            \\  
BLG217          & 002.7$-$03.7 & PHR 1806$-$2824 & 18 06 32.2 & $-$28 24 27 & P            \\  
BLG217,218 & 002.9$-$03.6 & MPA 1806$-$2812 & 18 06 39.3 & $-$28 12 21 & NULL          \\ 

BLG137          & 357.6$-$06.5 & PHR 1806$-$3416 & 18 06 46.3 & $-$34 16 04 & L             \\  
BLG235          & 003.9$-$03.1 & KFL 7 & 18 06 49.9 & $-$27 06 19          & NULL           \\  
BLG218          & 003.0$-$03.6 & MPA 1806$-$2807 & 18 06 52.9 & $-$28 07 05 & L          \\
BLG185          & 000.9$-$04.8 & M 3-23 & 18 07 06.2 & $-$30 34 17         & L            \\  
BLG160          & 359.4$-$05.6 & BMP 1807$-$3215 & 18 07 07.0 & $-$32 15 22 & L,B?            \\  
BLG197          & 002.1$-$04.2 & H 1-54 & 18 07 07.2 & $-$29 13 06         & NULL           \\
BLG197          & 001.7$-$04.4 & H 1-55 & 18 07 14.6 & $-$29 41 25         & NULL           \\
BLG235          & 003.8$-$03.2 & PHR 1807$-$2715 & 18 07 14.9 & $-$27 15 51 & L,B?             \\  
BLG168          & 359.9$-$05.4 & KFL 9 & 18 07 19.4 & $-$31 42 55          & L            \\  
BLG153          & 358.8$-$06.0 & MPA 1807$-$3254 & 18 07 22.0 & $-$32 54 31 & P           \\  
BLG197          & 001.7$-$04.6 & H 1-56 & 18 07 54.0 & $-$29 44 34         & T,S?           \\  
BLG242          & 004.4$-$03.1 & PHR 1807$-$2637 & 18 08 00.2 & $-$26 37 38 & P            \\  
BLG235          & 004.2$-$03.2 & KFL 10 & 18 08 01.4 & $-$26 54 02         & T            \\  
BLG218          & 002.9$-$04.0 & H 2-39 & 18 08 05.8 & $-$28 26 10         & L             \\  
BLG197          & 002.2$-$04.3 & PHR 1808$-$2913 & 18 08 06.5 & $-$29 13 12 & T            \\  
BLG169          & 359.7$-$05.7 & PHR 1808$-$3201 & 18 08 11.8 & $-$32 01 28 & T,B            \\  
BLG169          & 000.1$-$05.6 & H 2-40 & 18 08 30.7 & $-$31 36 36         & L            \\  
BLG161          & 359.3$-$06.0 & SB 54 & 18 08 31.4 & $-$32 29 59          & L,B?            \\  
BLG177          & 000.5$-$05.3 & SB 2 & 18 08 35.0 & $-$31 06 51           & L            \\  
BLG250          & 005.1$-$03.0 & H 1-58 & 18 09 13.9 & $-$26 02 29         & T,S            \\  
BLG161          & 359.4$-$06.3 & PPA 1809$-$3233 & 18 09 52.8 & $-$32 33 44 & L           \\
BLG236          & 004.1$-$03.8 & KFL 11 & 18 10 12.2 & $-$27 16 35         & NULL           \\
BLG236          & 004.4$-$03.8 & PPA 1810$-$2700 & 18 10 30.1 & $-$27 00 09 & P            \\  
BLG219          & 003.2$-$04.4 & KFL 12 & 18 10 30.7 & $-$28 19 23         & T            \\  
BLG219          & 003.3$-$04.4 & PPA 1810$-$2813 & 18 10 36.0 & $-$28 13 47 & L           \\
BLG187          & 001.2$-$05.6 & PHR 1811$-$3042 & 18 11 02.7 & $-$30 42 12 & T            \\  
BLG187          & 001.3$-$05.6 & SB 5 & 18 11 15.4 & $-$30 37 50           & L,B?            \\  
BLG251          & 005.4$-$03.4 & PHR 1811$-$2557 & 18 11 20.4 & $-$25 57 35 & T,B            \\  
BLG219          & 003.3$-$04.6 & Ap 1-12 & 18 11 35.0 & $-$28 22 38        & T            \\  
BLG220          & 003.5$-$04.6 & NGC 6565 & 18 11 52.6 & $-$28 10 43       & NEB            \\  
BLG178          & 000.7$-$06.1 & SB 3 & 18 12 14.4 & $-$31 19 59            & T,B             \\  
BLG243          & 005.0$-$03.9 & H 2-42 & 18 12 23.0 & $-$26 32 54         & L            \\  
BLG252          & 005.7$-$03.6 & KFL 13 & 18 12 44.6 & $-$25 44 19         & L             \\  
BLG252          & 005.5$-$04.0 & H 2-44 & 18 13 40.6 & $-$26 08 39         & L            \\  
BLG265          & 006.8$-$03.4 & H 2-45 & 18 14 28.8 & $-$24 43 38         & NULL           \\
BLG222          & 004.2$-$05.9 & M 2-37 & 18 18 38.4 & $-$28 07 58         & L \\
\end{supertabular}

\end{center}

\end{appendix}
\end{document}